\documentclass[12pt]{article}

\usepackage{cite}
\usepackage{subfigure}
\usepackage{multirow}
\usepackage{helvet}
\usepackage{amsmath}
\usepackage{amssymb}
\usepackage{setspace}
\usepackage{setspace}
\usepackage[dvips]{graphicx}
\usepackage{epsfig}
\usepackage{empheq}
\usepackage{subfigure}

\begin{document}

\titlepage                                                    
 \begin{flushright}                                                    
 IPPP/15/34  \\
 DCPT/15/68 \\                                                                                                       
  \end{flushright} 
 \vspace*{0.5cm}
\begin{center}                                                    
{\Large \bf Exclusive physics at the LHC with SuperChic 2}\\

\vspace*{1cm}
                                                   
L.A. Harland--Lang$^{1}$, V.A. Khoze$^{2,3}$, M.G. Ryskin$^{3}$ \\                                                 
                                                   
\vspace*{0.5cm}
${}^1$Department of Physics and Astronomy, University College London, WC1E 6BT, UK   \\                                                    
${}^2$Institute for Particle Physics Phenomenology, University of Durham, Durham, DH1 3LE          \\
${}^3$Petersburg Nuclear Physics Institute, NRC Kurchatov Institute, Gatchina, \linebreak[4]St. Petersburg, 188300, Russia                                              
                                                    
\vspace*{1cm}                         

\begin{abstract}

\noindent We present a range of physics results for central exclusive production processes at the LHC, using the new \texttt{SuperChic 2} Monte Carlo event generator. This includes significant theoretical improvements and updates, most importantly a fully differential treatment of the soft survival factor, as well as a greater number of generated processes. We provide an overview of the latest theoretical framework, and consider in detail a selection of final states, namely exclusive 2 and 3 jets, photoproduced vector mesons, two--photon initiated muon and $W$ boson pairs and heavy $\chi_{c,b}$ quarkonia. 

\end{abstract}
                                   
\end{center}  

\section{Introduction}

Central Exclusive Production (CEP) is the reaction
\begin{equation}\nonumber
pp({\bar p}) \to p+X+p({\bar p})\;,
\end{equation}
where `$+$' signs are used to denote the presence of large rapidity gaps, separating the system $X$ from the intact outgoing protons (anti--protons). Over the last decade there has been a steady rise of theoretical and experimental interest in studies of this process in high--energy hadronic collisions, see~\cite{Albrow:2010yb,Tasevsky:2014cpa,Harland-Lang:2014lxa,Harland-Lang:2014dta} for reviews. Theoretically, the study of CEP requires the development of a framework which is quite different from that used to describe the inclusive processes more commonly considered at hadron colliders.  Moreover, the dynamics of the CEP process leads to unique predictions and effects which are not seen in the inclusive mode. Experimentally, CEP represents a very clean signal, with just the object $X$ and no other hadronic activity seen in the central detector (in the absence of pile up). 

The CEP process requires the $t$--channel exchange of a color--singlet object, so that the outgoing protons can remain intact. One possibility to achieve this is the two--photon fusion process $\gamma\gamma \to X$, where the radiated  photons couple to the electromagnetic charge of the whole protons. Alternatively, the process may be mediated purely by the strong interaction: provided the object X mass is large enough, this can be considered in the framework of pQCD, via the so--called Durham model~\cite{Harland-Lang:2014lxa,Harland-Lang:2014dta}. Finally it is possible for `photoproduction'  reactions to occur, where one proton interacts electromagnetically and one interacts strongly.

In any detailed phenomenological study of such processes, it is important to have a Monte Carlo (MC) implementation, so that theoretical predictions can be compared more directly with experimental measurements. For this reason the authors have previously produced the publicly available \texttt{SuperChic} MC~\cite{HarlandLang:2009qe,HarlandLang:2010ep}, for the CEP of lighter Standard Model (SM) objects within the Durham model. While first considering $\chi_{c,b}$ and $\eta_{c,b}$ quarkonia, this has subsequently been extended to include $\gamma\gamma$ and light meson pair ($\pi\pi$,$\eta(')\eta(')$...) production, as well as the photoproduction of C--odd vector mesons. Such processes have been measured and compared to the MC predictions at the Tevatron and LHC, see for example~\cite{Aaltonen:2011hi,LHCb,Aaij:2014iea}, with results that are generally in good agreement.  Other related available generators are the \texttt{ExHuME}~\cite{Monk:2005ji} and FPMC~\cite{Boonekamp:2011ky} MCs.

However, there exist a wider range of processes that are not included in earlier versions of \texttt{SuperChic}, but which have much phenomenological relevance, in particular in the light of the measurement possibilities for exclusive processes during Run--II of the LHC~\cite{yp}. Here, exclusive events may be measured with both protons tagged using the approved and installed AFP~\cite{CERN-LHCC-2011-012} and CT--PPS~\cite{Albrow:1753795} forward proton spectrometers, associated with the ATLAS and CMS central detectors, respectively, see also~\cite{Royon:2015tfa}, as well as using rapidity gap vetoes to select a dominantly exclusive event sample. This latter possibility is in particular relevant at LHCb, for which the relatively low instantaneous luminosity and wide rapidity coverage allowed by the newly installed HERSCHEL forward detectors~\cite{Albrow:2014lta} are highly favourable, while similar scintillation counters are also installed at ALICE~\cite{Schicker:2014wvk}.

Particularly relevant is the case of exclusive jet production, which has been observed by both CDF~~\cite{Aaltonen:2007hs} and D0~\cite{Abazov:2010bk} at the Tevatron, and for which already a sample of `exclusive--like' 2 and 3--jet events has been collected in a combined CMS+TOTEM run at 8 TeV~\cite{Osterberg:2014mta,CMS-DP-2013-004}; exclusive 3--jet production in particular has not been implemented in any public MC. A further topical process is the exclusive production of quarkonia pairs, measured by LHCb in~\cite{Aaij:2014rms} and considered theoretically in~\cite{Harland-Lang:2014efa}, but for which a MC implementation has previously not been made publicly available. 

In addition to including a wider range of processes such as these, there are a number of theoretical updates and improvements which it is important to consider. Most significantly, in all exclusive processes it is necessary to account for the probability that no additional particles are produced by soft proton--proton interactions, independent of the hard process: the so--called `survival factor'. In previous versions of \texttt{SuperChic} as well as in other generators~\cite{Monk:2005ji,Boonekamp:2011ky}  this is simply treated as a constant probability which suppresses the overall cross section. However, it is well known that the survival factor can depend sensitively on the final--state particle momenta, and so such an averaging will omit the influence this can have on the predicted distributions as well as only providing an approximate estimate of the (process--dependent) overall suppression.

With these considerations in mind, we present in this paper results of the new \texttt{SuperChic 2} MC generator. This contains a range of theoretical improvements compared to the previous version, most significantly including a fully differential treatment of the survival factor, maintaining the explicit dependence of this on the particle momenta in all cases. As well as the processes generated in the original MC, exclusive 2 and 3 jet, quarkonia ($J/\psi$ and $\psi(2S)$) pair, SM Higgs boson production and the photoproduction of $\rho$ and $\phi$ mesons are now implemented. In addition, the two--photon production of $\gamma\gamma$, $W^+W^-$ and lepton pairs are included; this is the first MC implementation of such photon--induces processes which includes a complete treatment of soft survival effects. The case of a $e^+e^-$ initial state is also implemented for these processes.

The outline of this paper is as follows. In Section~\ref{durt} we summarise the theoretical ingredients of the Durham model of QCD--mediated CEP, and describe how the soft survival factor can be included differentially in theoretical predictions and in a MC. In Section~\ref{sec:photo} we describe the theory of photon--induced processes, again providing details of how a full treatment of the survival factor can be achieved. In Sections~\ref{sec:excjets} to \ref{sec:quark} we present results of this MC for a range of processes: exclusive 2 and 3 jet production in Section~\ref{sec:excjets}; exclusive vector meson photoproduction in Section~\ref{sec:exvec}; two--photon induced $W^+W^-$ and lepton pair production in Section~\ref{sec:2photo}; heavy $\chi_{c,b}$ quarkonia production in Section~\ref{sec:quark}. In Section~\ref{sec:other} a summary is presented of all processes that are generated, including some motivation for specific measurements that may be performed at the LHC. In Section~\ref{sec:MC} we briefly describe the \texttt{SuperChic 2} MC and its public availability. Finally, in Section~\ref{conc} we present a summary and outlook.

\section{QCD processes}\label{durt}

\subsection{Basic formalism}
\begin{figure}
\begin{center}
\includegraphics[scale=1.2]{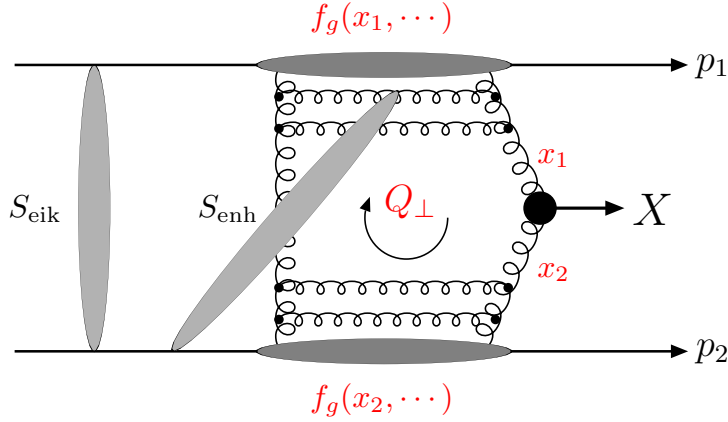}
\caption{The perturbative mechanism for the QCD--induced exclusive process $pp \to p\,+\, X \, +\, p$, with the eikonal and enhanced survival factors 
shown symbolically.}
\label{fig:pCp}
\end{center}
\end{figure} 
CEP processes that proceed purely by the strong interaction can be described by the `Durham' model, a pQCD--based approach that may be applied when the object mass $M_X$ is sufficiently high, see~\cite{Albrow:2010yb,Harland-Lang:2014lxa,Harland-Lang:2015eqa} for reviews. The formalism used to calculate the perturbative CEP cross section is explained in detail elsewhere~\cite{Khoze97,Khoze00,Kaidalov:2003fw,HarlandLang:2010ep,Khoze:2001xm,Albrow:2010yb,Harland-Lang:2013xba} and we will only present a very brief summary here. The perturbative CEP amplitude, corresponding to the diagram shown in Fig.~\ref{fig:pCp}, can be written as
\begin{equation}\label{bt}
T=\pi^2 \int \frac{d^2 {\bf Q}_\perp\, \overline{\mathcal{M}}}{{\bf Q}_\perp^2 ({\bf Q}_\perp-{\bf p}_{1_\perp})^2({\bf Q}_\perp+{\bf p}_{2_\perp})^2}\,f_g(x_1,x_1', Q_1^2,\mu_F^2; t_1)f_g(x_2,x_2',Q_2^2,\mu_F^2;t_2)\; ,
\end{equation}
where  $Q_\perp$ is the transverse momentum in the gluon loop, with the scale $Q_i^2 = Q_\perp^2$ in the forward proton limit (see e.g.~\cite{HarlandLang:2010ep} for a prescription away from this limit), and $\overline{\mathcal{M}}$ is the colour--averaged, normalised sub--amplitude for the $gg \to X$ process
\begin{equation}\label{Vnorm}
\overline{\mathcal{M}}\equiv \frac{2}{M_X^2}\frac{1}{N_C^2-1}\sum_{a,b}\delta^{ab}q_{1_\perp}^\mu q_{2_\perp}^\nu V_{\mu\nu}^{ab} \; .
\end{equation}
Here $a$ and $b$ are colour indices, $M_X$ is the central object mass, $V_{\mu\nu}^{ab}$ is the $gg \to X$ vertex, $q_{i_\perp}$ are the transverse momenta of the incoming gluons, and $t_i$ is the squared momentum transfer to the outgoing protons. The $f_g$'s in (\ref{bt}) are the skewed unintegrated gluon densities of the proton. These correspond to the distribution of gluons in transverse momentum $Q_\perp$, which are evolved in energy up to the hard scale $\mu_F$, such that they are accompanied by no additional radiation, as is essential for exclusive production. While the gluon momentum fractions $x_i$ are set by the mass and rapidity of the final state, the fractions ${x_i}'$ carried by the screening gluon must in general be integrated over at the amplitude level. However, for the dominant imaginary part of the amplitude we have $x' \ll x$, and it can be shown that the $f_g$'s may be simply written as
\begin{equation}\label{fgskew}
f_g(x,x',Q_\perp^2,\mu_F^2) =\; \frac{\partial}{\partial \ln(Q_\perp^2)} \left[ H_g\left(\frac{x}{2},\frac{x}{2};Q_\perp^2\right) \sqrt{T_g(Q_\perp,\mu_F^2)} \right]\;,
\end{equation}
where $H_g$ is the generalised gluon PDF~\cite{Belitsky:2005qn}, which for CEP kinematics can be related to the conventional PDFs~\cite{Shuvaev:1999ce,Harland-Lang:2013xba}. The $T_g$ in (\ref{fgskew}) is a Sudakov factor, which corresponds to the probability of no extra parton emission from each fusing gluon. 

We can decompose (\ref{Vnorm}) in terms of on--shell helicity amplitudes, neglecting small off--shell corrections of order $\sim {\bf q}_\perp^2/M_X^2$. Omitting colour indices for simplicity, this gives
\begin{align}
q_{1_\perp}^i q_{2_\perp}^j V_{ij} =\begin{cases} &-\frac{1}{2} ({\bf q}_{1_\perp}\cdot {\bf q}_{2_\perp})(T_{++}+T_{--})\;\;(J^P_z=0^+)\\ 
&-\frac{i}{2} |({\bf q}_{1_\perp}\times {\bf q}_{2_\perp})|(T_{++}-T_{--})\;\;(J^P_z=0^-)\\ 
&+\frac{1}{2}((q_{1_\perp}^x q_{2_\perp}^x-q_{1_\perp}^y q_{2_\perp}^y)+i(q_{1_\perp}^x q_{2_\perp}^y+q_{1_\perp}^y q_{2_\perp}^x))T_{-+}\;\;(J^P_z=+2^+)\\ 
&+\frac{1}{2}((q_{1_\perp}^x q_{2_\perp}^x-q_{1_\perp}^y q_{2_\perp}^y)-i(q_{1_\perp}^x q_{2_\perp}^y+q_{1_\perp}^y q_{2_\perp}^x))T_{+-}\;\;(J^P_z=-2^+)
\end{cases}\label{Agen}
\end{align}
where the $J_z^P$ indicate the parity and spin projection on the $gg$ axis, and $T_{\lambda_1\lambda_2}$ are the corresponding $g(\lambda_1)g(\lambda_2)\to X$ helicity amplitudes, see~\cite{HarlandLang:2010ep,Harland-Lang:2014lxa} for more details. In the forward proton limit (i.e. with outgoing proton $p_\perp=0$)  the only non-vanishing term after the $Q_\perp$ integration (\ref{bt}) is the first one: this is the origin of the selection rule~\cite{Kaidalov:2003fw,Khoze:2000mw,Khoze:2000jm} which operates in this exclusive process, and strongly favours $J^P_z=0^+$ quantum numbers for the centrally produced state. More generally, away from the exact forward limit the non-$J^P_z=0^+$ terms in (\ref{Agen}) do not give completely vanishing contributions to the $Q_\perp$ integral and we find that 
\begin{equation}\label{simjz2}
\frac{|A(|J_z|=2)|^2}{|A(J_z=0)|^2} \sim \frac{\langle p_\perp^2 \rangle^2}{\langle Q_\perp^2\rangle^2}\;,
\end{equation}
 which is typically of order $\sim1/50-1/100$, depending on such factors as the central object mass, c.m.s. energy $\sqrt{s}$ and choice of PDF set~\cite{HarlandLang:2010ep,Harland-Lang:2014lxa}. The on--shell decomposition (\ref{Agen}) is used throughout, unless otherwise stated.

\subsection{Soft survival effects}\label{sec:survqcd}
The expression (\ref{bt}) corresponds to the amplitude for the exclusive production of an object $X$ in a short--distance interaction, that is, with no further perturbative emission. However secondary particles may also be produced by additional soft proton--proton interactions, independent of the hard process. Such underlying event activity will spoil the exclusivity of the event, and the probability that no additional particles are produced by accompanying soft proton--proton interactions is encoded in the so--called `survival factor', see e.g.~\cite{Khoze:2014aca,Gotsman:2014pwa} for some more recent theoretical work, and~\cite{Harland-Lang:2014lxa} for further discussion and references.

The survival factor is not a simple multiplicative constant~\cite{HarlandLang:2010ep}, but rather depends quite sensitively on the outgoing proton transverse momenta. Physically, this is to be expected, as the survival factor will  depend on the impact parameter of the colliding protons; loosely speaking, as the protons become more separated in impact parameter, we should expect there to be less additional particle production, and so for the survival factor to be closer to unity (consequently, as we will see below, the average survival factor is much larger in the case of photon--mediated processes, where larger impact parameters are favoured, when compared to purely QCD processes). As the transverse momenta ${\bf p}_{i_\perp}$ of the scattered protons are nothing other than the Fourier conjugates of the proton impact parameters, ${\bf b}_{it}$, we therefore expect the survival factor to depend on these.

For this reason, survival effects are included fully differentially in the final--state momenta in \texttt{SuperChic 2}. To describe in more detail how this is achieved, we can consider a simplified `one--channel' model, which ignores any internal structure of the proton; see ~\cite{Ryskin:2009tj,Ryskin:2011qe} for discussion of how this can be generalised to the more realistic `mutli--channel' case. The average suppression factor is written as
\begin{equation}\label{S2}
\langle S^2_{\rm eik} \rangle=\frac{\int {\rm d}^2 {\bf b}_{1t}\,{\rm d}^2 {\bf b}_{2t}\, |T(s,{\bf b}_{1t},{\bf b}_{2t})|^2\,{\rm exp}(-\Omega(s,b_t))}{\int {\rm d}^2\, {\bf b}_{1t}{\rm d}^2 {\bf b}
_{2t}\, |T(s,{\bf b}_{1t},{\bf b}_{2t})|^2}\;,
\end{equation}
where ${\bf b}_{it}$ is the impact parameter vector of proton $i$, so that ${\bf b}_t={\bf b}_{1t}+{\bf b}_{2t}$ corresponds to the transverse separation between the colliding protons, with $b_t = |{\bf b}_t|$.  $T(s,{\bf b}_{1t},{\bf b}_{2t})$ is the CEP amplitude (\ref{bt}) in impact parameter space, and $\Omega(s,b_t)$ is the proton opacity, which can be extracted from such hadronic observables as the elastic and total cross sections as well as, combined with some additional physical assumption about the composition of the proton, the single and double diffractive cross sections. From (\ref{S2}), we can see that physically $\exp(-\Omega(s,b_t))$ represents the probability that no inelastic scattering occurs at impact parameter $b_t$. 

In the expression above, $T(s,{\bf b}_{1t},{\bf b}_{2t})$ is just the Fourier conjugate of the CEP amplitude (\ref{bt}), i.e. we have
\begin{equation}\label{Mfor}
T(s,{\bf p}_{1_\perp},{\bf p}_{2_\perp})=\int {\rm d}^2{\bf b}_{1t}\,{\rm d}^2{\bf b}_{2t}\,e^{i{\bf p}_{1_\perp}\cdot {\bf b}_{1t}}e^{-i{\bf p}_{2_\perp}\cdot {\bf b}_{2t}}T(s,{\bf b}_{1t},{\bf b}_{2t})\;.
\end{equation}
In transverse momentum space, the CEP amplitude including rescattering effects, $T^{\rm res}$, is calculated by integrating over the transverse momentum ${\bf k}_\perp$ carried round the Pomeron loop (represented by the grey oval labeled `$S_{\rm eik}^2$' in Fig.~\ref{fig:pCp}). The amplitude including rescattering corrections is given by
\begin{equation}\label{skt}
T^{\rm res}(s,\mathbf{p}_{1_\perp},\mathbf{p}_{2_\perp}) = \frac{i}{s} \int\frac{{\rm d}^2 \mathbf {k}_\perp}{8\pi^2} \;T_{\rm el}(s,{\bf k}_\perp^2) \;T(s,\mathbf{p'}_{1_\perp},\mathbf{p'}_{2_\perp})\;,
\end{equation}
where $\mathbf{p'}_{1_\perp}=({\bf p}_{1_\perp}-{\bf k}_\perp)$ and $\mathbf{p'}_{2_\perp}=({\bf p}_{2_\perp}+{\bf k}_\perp)$, while $T^{\rm el}(s,{\bf k}_\perp^2)$ is the elastic $pp$ scattering amplitude in transverse momentum space, which is related to the proton opacity via
\begin{equation}\label{sTel}
T_{\rm el}(s,t)=2s \int {\rm d}^2 {\bf b}_t \,e^{i{\bf k} \cdot {\bf b}_t} \,T_{\rm el}(s,b_t)=2is \int {\rm d}^2 {\bf b}_t \,e^{i{\bf k} \cdot {\bf b}_t} \,\left(1-e^{-\Omega(s,b_t)/2}\right)\;,
\end{equation}
where $t=-{\bf k}_\perp^2$. We must add (\ref{skt}) to the `bare' amplitude excluding rescattering effects to give the full amplitude, which we can square to give the CEP cross section including eikonal survival effects
\begin{equation}\label{Tphys}
\frac{{\rm d}\sigma}{{\rm d}^2\mathbf{p}_{1_\perp} {\rm d}^2\mathbf{p}_{2_\perp}} \propto |T(s,\mathbf{p}_{1_\perp},\mathbf{p}_{2_\perp})+T^{\rm res}(s,\mathbf{p}_{1_\perp},\mathbf{p}_{2_\perp})|^2 \;,
\end{equation}
where here (and above) we have omitted the dependence of the cross section on all other kinematic variables for simplicity. The overall normalisation of the cross section is achieved exactly as in the unscreened case. It is this expression, suitably generalised to the multi--channel case, which is used in the MC. We note that following the discussion above, the expected soft suppression can be written in transverse momentum space as
\begin{equation}\label{seikav1}
\langle S_{\rm eik}^2\rangle= \frac{\int {\rm d}^2{\bf p}_{1_\perp}\,{\rm d}^2{\bf p}_{2_\perp}\,|T(s,\mathbf{p}_{1_\perp},\mathbf{p}_{2_\perp})+T^{\rm res}(s,\mathbf{p}_{1_\perp},\mathbf{p}_{2_\perp})|^2}{\int {\rm d}^2{\bf p}_{1_\perp}\,{\rm d}^2{\bf p}_{2_\perp}\,|T(s,\mathbf{p}_{1_\perp},\mathbf{p}_{2_\perp})|^2}\;.
\end{equation}
It can readily be shown that (\ref{S2}) and (\ref{seikav1}) are equivalent. As expected, the soft suppression factor depends on the proton transverse momenta, and so may have an important effect on the distributions of the outgoing proton momenta, via (\ref{Tphys}), see e.g.~\cite{Khoze:2002nf,Harland-Lang:2013dia}.

Besides the effect of eikonal screening $S_{\rm eik}$, there is some suppression caused by the rescatterings of the protons with the intermediate partons~\cite{Ryskin:2009tk,Martin:2009ku,Ryskin:2011qe}. This effect is described by the so-called enhanced Reggeon diagrams and usually denoted as $S^2_{\rm enh}$, see Fig.~\ref{fig:pCp}. The precise size of this effect is uncertain, but due to the relatively large transverse momentum (and so smaller absorptive cross section $\sigma^{\rm abs}$) of the intermediate partons, it is only expected to reduce the corresponding CEP cross section by a factor of at most a `few', that is a much weaker suppression than in the case of the eikonal survival factor. Due to this uncertainty, in the current version of the MC these effects are omitted entirely, however by observing any departure from the MC predictions, for example in the invariant mass $M_X$ distributions, such enhanced survival effects may still be investigated.

Finally, it is worth pointing out that such soft survival effects do not only manifest themselves in CEP reactions. For example, the eikonal model of absorption discussed above may be used to predict the behaviour of the leading neutron spectra in diffractive dijet photoproduction, measured using a leading proton detector at HERA, see e.g.~\cite{Khoze:2006hw,Kaidalov:2009fp,Andreev:2015cwa}. In addition, this approach may be used to explain~\cite{Kaidalov:2001iz} the breaking of factorisation for diffractive dijet production at the Tevatron~\cite{Affolder:2000vb}, when this is given in terms of diffractive structure functions measured at HERA. As we only consider CEP reactions in this paper, we will not discuss these possibilities further here.

\section{Photon mediated processes}\label{sec:photo}

\subsection{Basic formalism}\label{sec:photobare}
Exclusive photon--exchange processes in $pp$ collisions are described in terms of the equivalent photon approximation~\cite{Budnev:1974de}. The quasi--real photons are emitted by the incoming proton $i=1,2$ with a flux given by
\begin{equation}\label{WWflux}
{\rm d}N^T(\xi_i)=\frac{\alpha}{\pi}\frac{{\rm d}^2q_{i_\perp} }{q_{i_\perp}^2+\xi_i^2 m_p^2}\frac{{\rm d}\xi_i}{\xi_i}\left(\frac{q_{i_\perp}^2}{q_{i_\perp}^2+\xi_i^2 m_p^2}(1-\xi_i)F_E(Q_i^2)+\frac{\xi_i^2}{2}F_M(Q_i^2)\right)\;,
\end{equation}
where $\xi_i$  and $q_{i_\perp}$ are the longitudinal momentum fraction and transverse momentum of the photon $i$, respectively; in the absence of rescattering, we have simply $q_{i_\perp}=-p_{i_\perp}$, see Fig.~\ref{fig:pVp} below. The functions $F_E$ and $F_M$ are given in terms of the proton electric and magnetic form factors, via
\begin{equation}
F_M(Q^2_i)=G_M^2(Q^2_i)\qquad F_E(Q^2_i)=\frac{4m_p^2 G_E^2(Q_i^2)+Q^2_i G_M^2(Q_i^2)}{4m_p^2+Q^2_i}\;,
\end{equation}
with
\begin{equation}
G_E^2(Q_i^2)=\frac{G_M^2(Q_i^2)}{7.78}=\frac{1}{\left(1+Q^2_i/0.71 {\rm GeV}^2\right)^4}\;,
\end{equation}
in the dipole approximation, where $G_E$ and $G_M$ are the `Sachs' form factors. The modulus of the photon virtuality, $Q^2_i$, is given by
\begin{equation}\label{qi}
Q^2_i=\frac{q_{i_\perp}^2+\xi_i^2 m_p^2}{1-\xi_i}\;,
\end{equation}
 i.e. it is cut off at a kinematic minimum $Q^2_{i,{\rm min}}=\xi_i^2 m_p^2/(1-\xi_i)$.

 The cross section for the photoproduction of a state $V$, for the case that the photon is emitted from proton $i$, is then simply given in terms of the photon flux (\ref{WWflux}) and the $\gamma p \to V p$ subprocess cross section
 \begin{equation}\label{sigphot}
 \sigma_{pp \to pVp}^i = \int {\rm d}N^T(\xi_i) \,\sigma_{\gamma p \to Vp}^i\;,
 \end{equation}
 integrated over the relevant phase space region. As the transverse momentum transferred by the photon exchange is typically much smaller than that due to the proton--Pomeron vertex, we may safely ignore interference effects, so that the total cross section is simply given by summing over $i=1,2$, i.e. allowing for the case that the photon is emitted from either proton. For two--photon production $\gamma\gamma \to X$, the corresponding cross section is
\begin{eqnarray}\label{siggam}
 \frac{{\rm d}\sigma_{pp\to pXp}}{{\rm d}\Omega}&=&\int\frac{{\rm
     d}\sigma_{\gamma\gamma\to X}(W_{\gamma\gamma})}{{\rm d}\Omega}\frac{{\rm
     d}L^{\gamma\gamma}}{{\rm d}W_{\gamma\gamma}}{\rm d W_{\gamma\gamma}}\;,
\end{eqnarray}
where $W_{\gamma\gamma}$ is the $\gamma\gamma$ c.m.s. energy. The $\gamma\gamma$ luminosity is  given by
\begin{equation}\label{lgam}
\frac{{\rm d}L^{\gamma\gamma}}{{\rm d}W_{\gamma\gamma}\,{\rm d}y_X}= \frac{2 W_{\gamma\gamma}}{s}\,n(x_1) \,n(x_2)\;,
\end{equation}
where $y_X$ is the object rapidity and $x_{1,2}=\frac{W_{\gamma\gamma}}{\sqrt{s}}\exp(\pm y_X)$, while $n(x_i)$ is the photon number density:
\begin{equation}
n(x_i)=\int  {\rm d}N^T(\xi_i) \,\delta(\xi_i - x_i)\;.
\end{equation}

\subsection{Soft survival effects}\label{sec:photoscreen}
For photon--mediated processes, survival effects can be included exactly as described in Section~\ref{sec:survqcd}, 
however some additional care is needed. From (\ref{Tphys}) we can see that it is the amplitude for the production process that is the relevant object when including these effects. On the other hand, (\ref{sigphot}) and (\ref{siggam}) and the flux (\ref{WWflux}) are defined at the cross section level, with the squared amplitude for the photon--initiated subprocesses summed over the (transverse) photon polarisations. 

\begin{figure}
\begin{center}
\subfigure[bare]{
\includegraphics[scale=0.88]{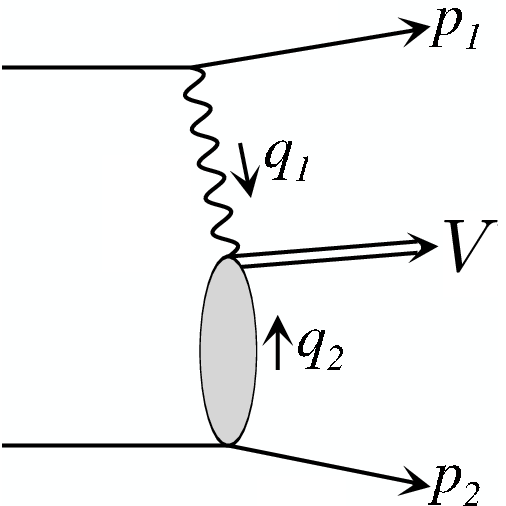}
}\qquad\qquad
\subfigure[screened]{
\includegraphics[scale=0.8,clip=true]{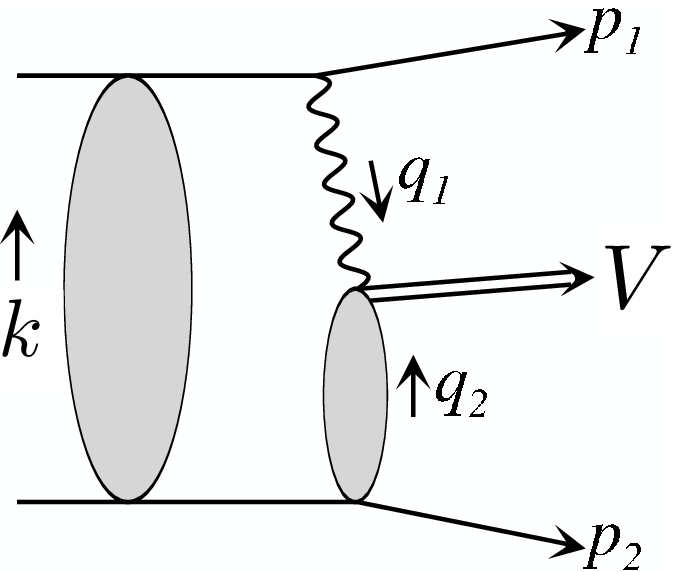}
}
\caption{Schematic diagrams for the exclusive photoproduction process $pp\to pVp$ with (a) and without (b) screening corrections included.}
\label{fig:pVp}
\end{center}
\end{figure} 

To translate these expressions to the appropriate amplitude level, it is important to include the photon transverse momentum $q_\perp$ dependence in the appropriate way, corresponding to a correct treatment of the  photon polarisation, see~\cite{Khoze:2002dc}. To demonstrate this, we will only consider the $F_E$ term in (\ref{WWflux}) in what follows, but will comment on the contribution of the magnetic form factor at the end. Schematic diagrams for the bare and screened photoproduction amplitudes are shown in Fig.~\ref{fig:pVp},  with the relevant momenta indicated; for the bare amplitude we have $q_{i_\perp}=-p_{i_\perp}$, while for the screened we have $q_{i_\perp}=-p'_{i_\perp}$, see Section~\ref{sec:survqcd}.  Using the same decomposition that leads to (\ref{Agen}), the photoproduction amplitude corresponding to the figure behaves as 
\begin{equation}\label{tgam}
  T(q_{1_\perp}) \sim q_{1_\perp}^x (A^+ - A^-)+i q_{1_\perp}^y (A^+ + A^-)\;,
 \end{equation} 
  where $A^\pm$ is the $\gamma p\to V p$ amplitude for a photon of $\pm$ helicity, and if the photon is emitted from the other proton we simply interchange $1\leftrightarrow 2$. In the bare case ($q_{1\perp}=-p_{1\perp}$)  we simply square this, and after performing the azimuthal angular integration, the cross terms $\sim p_{1_\perp}^x p_{1_\perp}^y$ vanish and we have
  \begin{equation}\label{tgamsq}
  |T(p_{1_\perp}) |^2 \sim  p_{1_\perp}^2 \sigma_{\gamma p \to Vp}\;,
  \end{equation}
  where $\sigma_{\gamma p \to Vp}$ is the subprocess cross section summed over the incoming photon transverse polarisations. This is consistent with the $F_E$ term in (\ref{WWflux}) and with (\ref{sigphot}), and indeed a full treatment, keeping all prefactors and expressing the $pp \to pVp$ cross section in terms of (\ref{tgamsq}), leads to exactly these results, and is the essence of the equivalent photon approximation.
  
When calculating the screened amplitude it is crucial to correctly include this explicit transverse momentum dependence as in (\ref{tgam}), with $q_{1_\perp} = -{p'}_{1_\perp}$ included inside the integral (\ref{skt}); this vector structure of the amplitude can have a significant effect on the expected survival factor. More precisely for the photoproduction amplitude we take
\begin{equation}
T(q_{1_\perp},q_{2\perp}) = T' \frac{F_E(Q_1^2)^{1/2}}{q_{1_\perp}^2+\xi_1^2 m_p^2}((q_{1_\perp}^x (A^+ - A^-)+i q_{1_\perp}^y (A^+ + A^-))\;,
\end{equation}
where $T'$ contains the transverse momentum dependence of the other proton (i.e $T' \sim e^{-b_V q_{2_\perp}^2/2}$ in the fit of (\ref{sigmaw})), as well as the $\xi$ dependence (and other factors) in (\ref{WWflux}) and the $\gamma p$ c.m.s. energy $W_{\gamma p}$ dependence of the $\gamma p \to V p$ subprocess.

For the case of two--photon initiated processes, the amplitude can be decomposed precisely as in (\ref{Agen}), i.e.
\begin{equation}
T(q_{1_\perp},q_{2\perp}) \sim -\frac{1}{2} ({\bf q}_{1_\perp}\cdot {\bf q}_{2_\perp})(T_{++}+T_{--})+\cdots\;,\label{Agenphot}
\end{equation}
where the $T_{\lambda_1\lambda_2}$ are now the $\gamma(\lambda_1)\gamma(\lambda_2)\to X$ helicity amplitudes, and we omit the overall factors for simplicity. Using this, it can readily be shown that the bare amplitude squared reduces to the correct cross section level expressions given in Section~\ref{sec:photobare}, while in the screened case it is again crucial to include this correct vector form of the amplitude, i.e. with the $q_{i_\perp}= -{p'}_{i_\perp}$ included inside the integral (\ref{skt}). As the relative contributions of the  amplitudes $T_{\lambda_1\lambda_2}$ affect the $q_\perp$ dependence in (\ref{Agenphot}), the survival factor may depend sensitively on the helicity structure of the $\gamma\gamma\to X$ process. For example, in the case of dilepton production $\gamma\gamma\to l^+ l^-$, for which the $T_{\pm \pm}$ amplitudes vanish for massless leptons, we find much less suppression than may be naively expected~\cite{Khoze:2000db}, see Section~\ref{sec:2photo} for further discussion.

Finally, we must also consider the contribution from the magnetic form factor $F_M$ in (\ref{WWflux}). While generally suppressed by $\xi^2$, for larger values of $M_X$ and/or production in the forward region, the corresponding value of $\xi$ may not be so small, and the contribution from this term may not be negligible. A careful consideration of the derivation of the equivalent photon approximation shows that this contribution is generated by a term $\sim g^{\mu\nu}$ given by the density matrix of the virtual photon (i.e. the proton spin sum) in the cross section. This is not proportional to $q_{i_\perp}^2$ and does not allow a decomposition, at the amplitude level, as in (\ref{tgam}); the $F_E$ contribution on the other hand is given by the term proportional to $q_{i_\perp}^\mu q_{i_\perp}^\nu$, as expected from (\ref{tgam}). Therefore, to evaluate the $F_M$ contribution we simply omit any such $q_{i_\perp}^\mu$ dependence when calculating the screened amplitude (\ref{skt}). For the photoproduction case, we then add this squared amplitude incoherently to the $F_E$ term, which is calculated as described above. For two--photon production, we keep the explicit vector $q_{i_\perp}$ dependence as in (\ref{Agenphot}) for the (dominant) $F_E(Q_1)F_E(Q_2)$ contribution, while for the other terms no explicit vector $q_{i_\perp}$ dependence is included in the amplitude, and the corresponding contributions are again squared and added incoherently.

\section{Physics processes}

In the following sections we consider a selection of  representative examples of the physics processes that are generated by \texttt{Superchic 2}. 

\subsection{Exclusive jet production}\label{sec:excjets}
\begin{table}
\begin{center}
\begin{tabular}{|c|c|c|c|c|c|}
\hline
$M_X({\rm min})$ & $gg$ & $q\overline{q}$ & $b\overline{b}$ & $ggg$ & $gq\overline{q}$ \\ \hline
75 & 120 & 0.073 & 0.12 &6.0 &0.14\\
150 &4.0  & $1.4 \times 10^{-3}$ &  $1.7 \times 10^{-3}$ &0.78 & 0.02\\
250 & 0.13 & $5.2 \times 10^{-5}$ &$5.2 \times 10^{-5}$   &0.018 &$5.0 \times 10^{-4}$ \\
\hline
\end{tabular}
\caption{Parton--level predictions for exclusive two and three jet production cross sections (in pb) at the LHC for different cuts on the minimum central system invariant mass $M_X$ at $\sqrt{s}=13$ TeV. The jets are required to have transverse momentum $p_\perp>$ 20 GeV for $M_X({\rm min})=75,150$ GeV and $p_\perp>$ 40 GeV for $M_X({\rm min})=250$ GeV and pseudorapidity $|\eta|<2.5$.  The anti--$k_t$ algorithm with jet radius $R=0.6$  is used in the three jet case and the $q\overline{q}$ cross sections correspond to one massless quark flavour. Soft survival effects are included using model 4 of~\cite{Khoze:2013dha}.}
\label{table:jets}
\end{center}
\end{table}
Exclusive jet production~\cite{Martin:1997kv,Khoze:2006iw}, in particular of a 2--jet system, has been of great importance in testing the underlying perturbative CEP formalism. As discussed in the introduction, it has been observed at the Tevatron~\cite{Aaltonen:2007hs,Abazov:2010bk}, and there is much potential to measure this process at the LHC, in particular with the protons tagged with forward proton spectrometers associated with the ATLAS and CMS central detectors, see~\cite{Royon:2015tfa}. Most events with two scattered protons and central jets will correspond to central diffractive (CD) jet production, i.e they will not be truly exclusive, but will have additional particle production from the Pomeron remnants. Exclusive production may be regarded as a particular case of CD jet production with only the jets in the final state, and no Pomeron remnants. It proceeds through the mechanism shown in Fig.~\ref{fig:pCp}, via the $gg\to gg, q\overline{q}$ and $gg \to ggg,gq\overline{q}$ subprocesses for 2-- and 3--jet production, respectively. 

\begin{figure}[h]
\begin{center}
\includegraphics[scale=0.6]{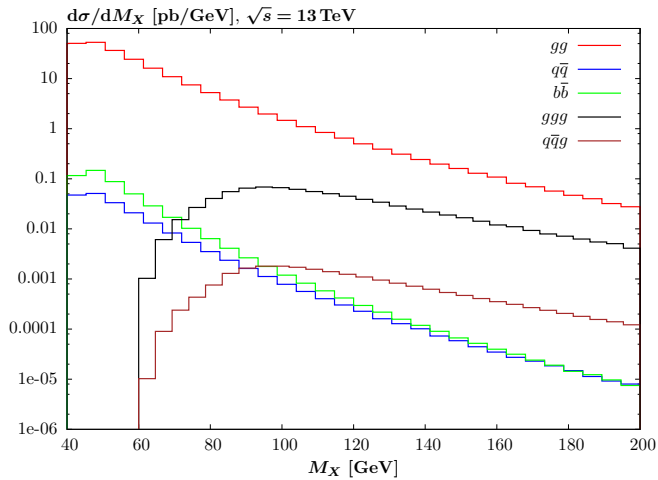}
\caption{Parton--level distributions for 2 and 3--jet CEP with respect to the system invariant mass $M_X$ at $\sqrt{s}=13$ TeV, using MMHT14 LO PDFs~\cite{Harland-Lang:2014zoa}. The final--state partons are required to lie in the pseudorapidity region $-2.5<\eta<2.5$ and have transverse momentum $p_\perp>20$ GeV (leading to minimum invariant masses, $M_X$, of 40 and 60 GeV in the two and three jet cases, respectively), while the three--jet events are defined using the anti--$k_t$ algorithm with $R=0.6$. Distributions are shown for massless quarks, $q\overline{q}$, as well as for $b\overline{b}$ production. Soft survival effects are included using model 4 of~\cite{Khoze:2013dha}.}
\label{fig:mjet}
\end{center}
\end{figure}

Further details about the contributing helicity amplitudes are given in Appendix~\ref{app:jets}. It is found, in particular, that in the case of the $gg\to q\overline{q}$ process, the $J_z=0$ amplitude (\ref{qqexcjz0}) involves a helicity flip along the quark line, and vanishes as the quark mass $m_q \to 0$. From this fact, we expect a strong suppression in the CEP cross section for quark dijets, relative to the $gg$ case, for which the $gg\to gg$ amplitudes (\ref{ggexcjz0}) with $J_z=0$ incoming gluons display no such suppression. In this way the exclusive mode offers the possibility to study almost purely (over $99\%$ for typical event selections) gluonic and, crucially, isolated jets~\cite{Khoze:2000jm} produced by the collision of a color--singlet $gg$ state, shedding light on their underlying properties (such as multiplicity, particle correlations etc) in a well--defined and comparatively clean exclusive environment. In Table~\ref{table:jets}, some representative predictions for exclusive two and three jet production are shown and this $gg/q\overline{q}$ hierarchy is clear. The invariant mass distributions are shown in Fig.~\ref{fig:mjet}, with similar results being evident.    
     
In the case of three jet production, that is $q\overline{q}g$ and $ggg$ jets, this suppression in the $q\overline{q}$ exclusive dijet cross section also leads to some interesting predictions~\cite{Khoze:2006um,Khoze:2009er}. In particular, we expect the behaviour of the $q\overline{q}g$ amplitude as the radiated gluon becomes soft to be governed by the corresponding Born--level, $q\overline{q}$, amplitude. As this vanishes for massless quarks and $J_z=0$ incoming gluons, it is expected to lead to an enhancement of `Mercedes--like' configurations for the $q\overline{q}g$ case, where all three partons carry roughly equal energies and are well separated. The corresponding three--jet cross sections are also shown in Table~\ref{table:jets}: while the $gg$ dijet cross sections are of order $\sim 100$ pb, the three--jet $ggg$ cross section are a factor of $\sim 10$ smaller, and the $q\overline{q}g$ cross section a further order of magnitude smaller again; this is due to the specific colour and spin--dependence of the contributing $gg\to q\overline{q}g$ amplitudes, which also leads to some suppression in the inclusive case, as well as the additional dynamical suppression discussed above. The corresponding invariant mass distributions are shown in Fig.~\ref{fig:mjet}.

\begin{figure}
\begin{center}
\includegraphics[scale=0.55]{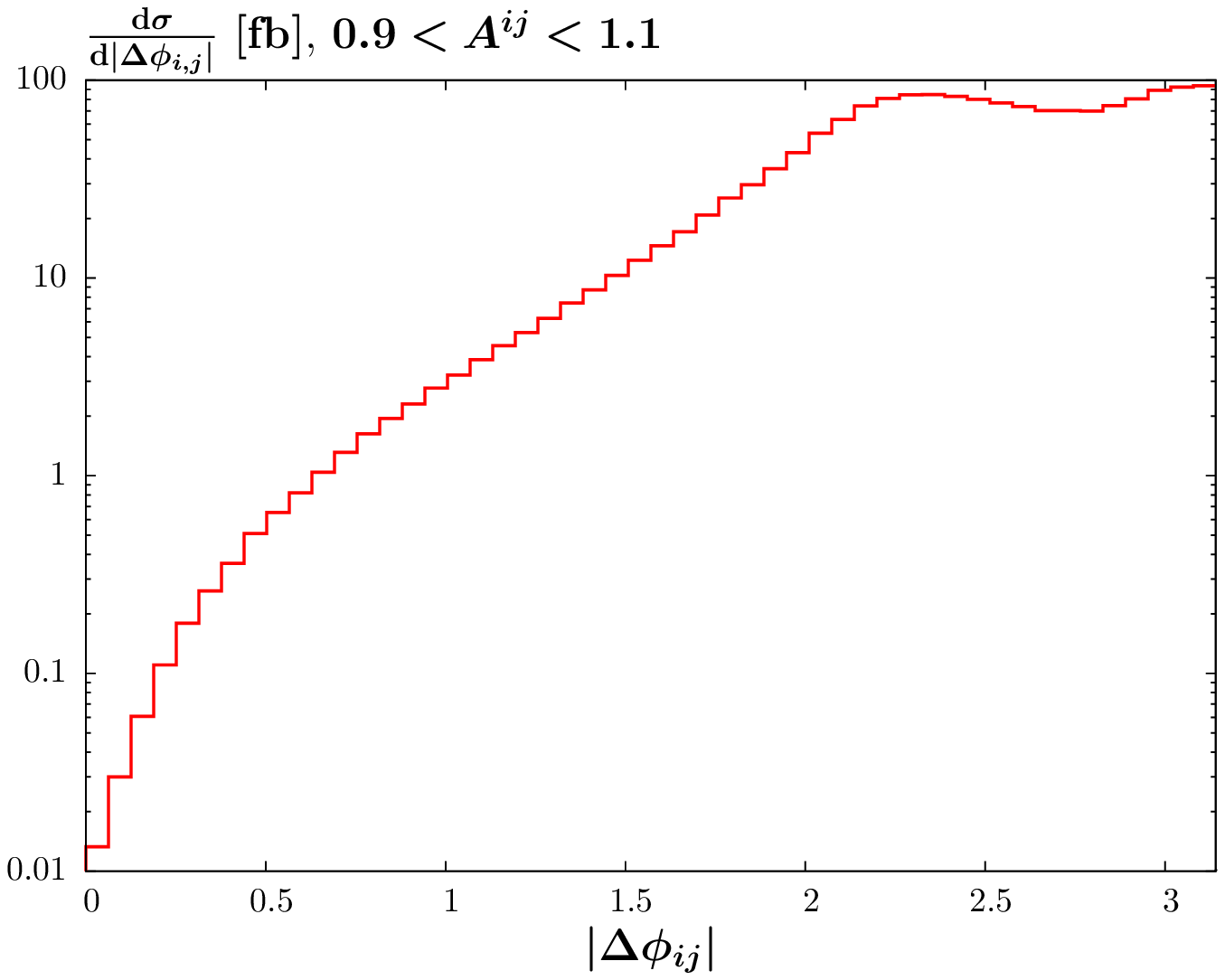}\quad
\includegraphics[scale=0.55]{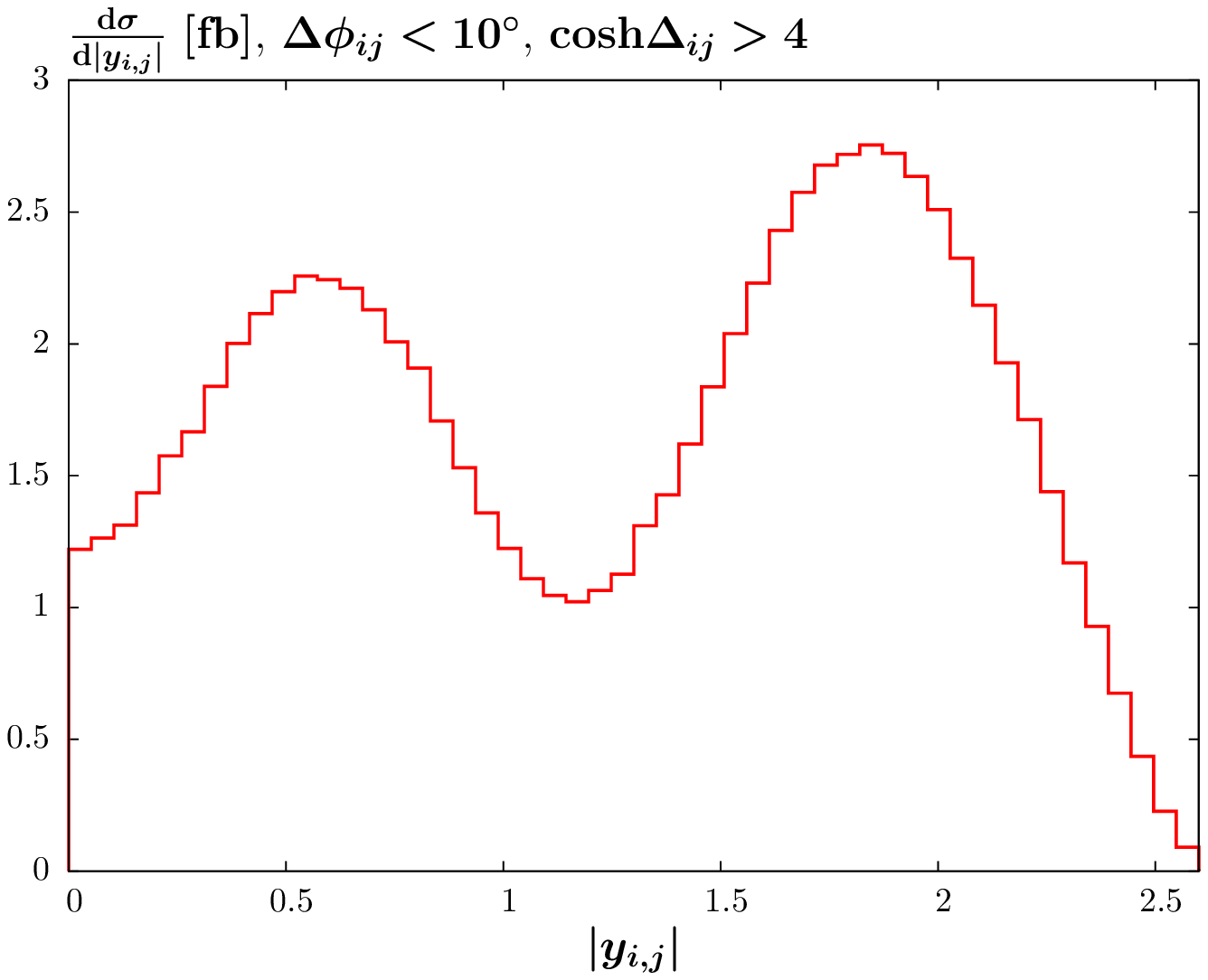}
\caption{Differential cross sections (in pb) at $\sqrt{s}=13$ TeV for exclusive $ggg$ jet production at parton level, with respect to (left) the separation in azimuthal angle  $\Delta\phi_{ij}$ for gluon pairs $ij$ passing the cut $0.9<A_{ij}<1.1$, where $A_{ij}$ is defined in (\ref{aij}), (right) the rapidity differences $|y_{i,j}-y_X|$ for which ${\rm cosh}\Delta_{ij}>4$ and $\Delta\phi_{ij}<10^\circ$, where $y_X$ is the rapidity of the 3--jet system and the $y_{i,j}$ is the rapidity of parton $i,j$, both of which are binned. All partons are required to have $p_\perp>20$ GeV and $|y|<5$.}
\label{fig:rad0a}
\end{center}
\end{figure}

So--called `planar radiation zeros'  were shown in~\cite{Harland-Lang:2015faa} to be present in 5--parton QCD amplitudes, that is, a complete vanishing of the Born--level amplitudes, independent of the particle polarisations, when their momenta lie in a plane and satisfy certain additional conditions on their rapidity differences. These were seen in particular to occur in the $gg \to ggg$ and, in certain cases, the $gg\to q\overline{q} g$ amplitudes, when the initial--state gluons are in a colour--singlet configuration. This is precisely the situation for exclusive 3--jet production, and so it is interesting to examine whether such zeros may be observable in the CEP process. In Fig.~\ref{fig:rad0a} (left) we show parton--level MC predictions for $ggg$ jet production with respect to the azimuthal separation  $\Delta\phi_{ij}$  for gluon pairs satisfying the cut $0.9<A_{ij}<1.1$, with
\begin{equation}\label{aij}
A_{ij}=\frac{{\rm sinh}^2\left(\frac{\Delta_{ij}}{2}\right)}{{\rm cosh}^2\left(\frac{\Delta_{jk}}{2}\right)+{\rm cosh}^2\left(\frac{\Delta_{ik}}{2}\right)}\;,
\end{equation}
where $i,j=1,2,3$ label the final--state gluons, and $\Delta_{ij}=y_i-y_j$. It was shown in~\cite{Harland-Lang:2015faa} that a zero occurs when a gluon pairing satisfies $A_{ij}=1$ and has zero azimuthal angular separation $\Delta\phi_{ij}=0$. 
In Fig.~\ref{fig:rad0a} (left), a clear suppression by several orders of magnitude for lower $\Delta\phi_{ij}$, which is driven by this zero, is evident; it was shown in~\cite{Harland-Lang:2015faa} that such a strong suppression is not seen inclusively. In Fig.~\ref{fig:rad0a} (right) we show predictions with respect to the rapidity difference $|y_{i,j}-y_X|$, with the cut  $\Delta\phi_{ij}<10^\circ$ applied on the angular separation, and with the additional requirement that ${\rm cosh}\Delta_{ij}>4$, which helps to isolate the region where $A_{ij}=1$ may be satisfied. Again a clear dip is seen in the distribution; as before it can be shown explicitly that this dip does not occur inclusively, and it is directly driven by the presence of a radiation zero, rather than being, say, an artefact of the cut choices.

While measurements of such distributions can in principle provide quite a clear demonstration of these radiation zeros, an important question is whether the expected signal size would be large enough. From Table~\ref{table:jets} we may expect $\sim 500$ signal $ggg$ events from $100\,{\rm pb}^{-1}$ of low pile--up running which can be realistically anticipated at the LHC at $\sqrt{s}=13$ TeV, for jet transverse momenta $p_\perp>25$ GeV, and rapidities $|\eta|<2.5$. By extending out to $|\eta|<5$ we may increase the event sample by a factor of $\sim 1.5$, and in addition it should be emphasised that the only necessary requirement of the production mechanism for these radiation zeros to be present is that the initial--state gluons be in a colour--singlet state. That is, the effect is independent of the particle polarisations, and of any additional kinematic effects specific to the pure CEP process, such as the $J_z^{PC}=0^{++}$ selection rule. Thus events where one or both protons dissociate, but with large rapidity gaps between the dissociation states and the jet system, for which the initial--state gluons are also in a colour--singlet configuration, will be expected to contain such zeros. Allowing for these, we may expect a sample of $\sim 1000$ $ggg$ events, while the $q\overline{q}g$ contribution is expected from Table~\ref{table:jets} to be $\sim$ an order of magnitude smaller, and will not be considered in what follows.

While such a sample may at first glance be sufficient to be sensitive to the type of dips shown in Fig.~\ref{fig:rad0a}, some caution is needed. We recall that we are interested in planar configurations of the jets, where the zeros may occur: after placing a cut of $\Delta\phi_{ij}<10^\circ(20^\circ)$ the expected sample is reduced to only 55 (142) events. Moreover, it can be shown that the zero condition $A_{ij}=1$ only has a solution if ${\rm cosh}\Delta_{ij}>7$, corresponding to $|\Delta_{ij}|\gtrsim 2.6$. Such a separation in rapidity, which leads to larger invariant masses of the three--jet system, is strongly suppressed, and placing such a cut on the original sample of 1000 events reduces it to only 36. A less restrictive cut may be placed as in Fig.~\ref{fig:rad0a} (right), of ${\rm cosh}\Delta_{ij}>4$, but after combining this with a reasonable cut on $\Delta\phi_{ij}$, less than 1 event remains. Clearly a measurement of a dip in such a rapidity distribution, even with a more fine--tuned choice of cuts, will be highly challenging during low--luminosity running at the LHC. 

\begin{table}[h]
\begin{center}
\renewcommand\arraystretch{1.4}
\begin{tabular}{|c|c|c|c|c|c|c|}
\cline{2-7}
\multicolumn{1}{}{}&\multicolumn{3}{|c|}{$0.7<A_{ij}<1.3$}&\multicolumn{3}{|c|}{$0.9<A_{ij}<1.1$}\\
\cline{2-7}
 \multicolumn{1}{c|}{}&$\Delta\phi_{ij}<60^\circ$ &$\Delta\phi_{ij}>60^\circ$ &$\frac{\Delta\phi_{ij}<60^\circ}{\Delta\phi_{ij}>60^\circ}$&$\Delta\phi_{ij}<60^\circ$ &$\Delta\phi_{ij}>60^\circ$&$\frac{\Delta\phi_{ij}<60^\circ}{\Delta\phi_{ij}>60^\circ}$  \\ \hline
Exclusive& 0.6 & 55& 1.1\%& 0.15&16&0.9\%\\
\hline
Inclusive& 10.3 & 157& 6.6\%&3.2&48&6.7\%\\
\hline
\end{tabular}
\caption{Expected number of $ggg$ jet events for which a gluon pairing $ij$ passes cuts on the azimuthal separation $\Delta\phi_{ij}$ and $A_{ij}$, defined in the text. Results for both exclusive and inclusive parton--level events, using the MMHT14LO PDFs~\cite{Harland-Lang:2014zoa} are shown, with an initial sample of 1000 events, before cuts, considered in both cases for illustration.}
\label{table:dphi}
\end{center}
\end{table}

On the other hand, if the cut $0.7(0.9)<A_{ij}<1.3(1.1)$ is placed on the same event sample, then 53(16) expected events remain, and even with such a fairly small number of events some discrimination may be possible by considering the azimuthal separation observable $\Delta\phi_{ij}$ as in Fig.~\ref{fig:rad0a} (left). In Table~\ref{table:dphi} the expected number of events for which a gluon pairing $ij$ passes such a cut on $A_{ij}$ and with $\Delta\phi_{ij}$ greater and less than $60^\circ$ is shown; for exclusive production, we anticipate a strong suppression in the $\Delta\phi_{ij}<60^\circ$ region. For comparison results for a colour summed inclusive sample of 1000 events are also shown: a strong  suppression in the ratio of events passing the $\Delta\phi_{ij}<60^\circ$ to $\Delta\phi_{ij}>60^\circ$ cuts in the exclusive case is seen in comparison to the inclusive. Although these results are at LO and parton--level only, and may be washed out somewhat in a more realistic treatment, including in particular parton shower/higher--order effects as well as background events, this may nonetheless be a promising measurement possibility. By comparing such a ratio with the measured sample of events dominantly due to CD jet production (for which there is no colour--singlet requirement and therefore no observable radiation zeros), a more robust signal of this suppression may be observable.

Finally, it is again worth emphasising that the only requirement on the production mechanism for such zeros to occur is that the initial--state gluons be in a colour--singlet state, with the particle polarisations playing no role, and so the pure CEP case is not the only possibility to observe these zeros. As described above, events with proton dissociation may also be considered, but it would also be interesting to examine if jet properties such as the colour flow and multiplicity might be used to isolate the contribution from a colour--singlet initial state, and so investigate these zeros in an inclusive environment. For example, the so--called jet pull angle variable has been shown to be useful in identifying jets which originate from a colour--singlet $gg$ initial state~\cite{CMS:2014joa,Aad:2015lxa}.

\subsection{Exclusive vector meson photoproduction}\label{sec:exvec}
In this section we will consider the photoproduction of vector mesons, focussing on the $J/\psi$ and $\Upsilon(1S)$ cases; although $\psi(2S)$ production is also included in the MC, it will not be considered here. At the LHC, coherent $J/\psi$ photoproduction in ultra peripheral $p$--$Pb$ collisions has been measured by ALICE at $\sqrt{s_{NN}}=5.02$ TeV~\cite{TheALICE:2014dwa}, and LHCb have made increasingly precise measurements of $J/\psi$ (and $\psi(2S)$) photoproduction in $pp$ collisions~\cite{Aaij:2013jxj,Aaij:2014iea} at $\sqrt{s}=7$ TeV. We will focus here on production (in $pp$ collisions) in the forward region relevant to the LHCb acceptance, but will also show some representative results for central production.

Following the notation of Fig.~\ref{fig:pVp}, for the $\gamma p \to V p$ subprocess cross section we take the power--law fit
\begin{equation}\label{sigmaw}
\frac{{\rm d}\sigma^{\gamma p \to V p}}{{\rm d} q_{2_\perp}^2}= N_V\left(\frac{W_{\gamma p}}{1 \,\rm GeV}\right)^{\delta_V} b_V\, e^{-b_V q_{2_\perp}^2}\;.
\end{equation}
For the case of $J/\psi$ production we take $N_\psi = 3.97$ {\rm nb} and $\delta_\psi = 0.64$, consistently with the HERA fit~\cite{Alexa:2013xxa}, which finds $N_\psi = 3.97\pm 0.05$ and $\delta_\psi=0.67\pm 0.03$; these precise choices will be justified below. For the $\Upsilon(1S)$ we take the values of $N_\Upsilon=0.12$ pb and $\delta_\Upsilon=1.6$ from~\cite{Motyka:2008ac}, although we note that in this case these are quite poorly constrained by the existing HERA data. The slope $b_V$ is fitted using a Regge--based parameterisation
\begin{equation}\label{bpsi}
b_V = b_0 + 4\alpha' \log \left(\frac{W_{\gamma p}}{90 \,{\rm GeV}}\right)\;,
\end{equation}
with $b_0 =4.6\,{\rm GeV}^{-2}$ and $\alpha'=0.2\,{\rm GeV}^{-2}$, consistently with the HERA measurement~\cite{Aktas:2005xu}. In the absence of any precise data in the cases of $\Upsilon(1S)$ and $\psi(2S)$ production, we assume that these values are universal.

\begin{table}[h]
\begin{center}
\renewcommand\arraystretch{1.15}
\begin{tabular}{|c|c|c|c|}
\cline{2-4}
\multicolumn{1}{}{}&\multicolumn{2}{|c|}{$2<\eta^\mu<4.5$}&\multicolumn{1}{|c|}{$-1<\eta^\mu<1$}\\
\cline{2-4}
 \multicolumn{1}{c|}{}& $\sqrt{s}=7$ TeV &  $\sqrt{s}=13$ TeV&  $\sqrt{s}=13$ TeV \\ \hline
$\sigma^{\psi}_{\rm bare} $& 359 & 511& 333 \\
$\sigma^{\psi}_{\rm sc.} $ & 278 & 406& 291  \\
$\langle S^2_{\rm eik}\rangle$ & 0.77 & 0.79& 0.87 \\
\hline
\end{tabular}
\caption{Cross section predictions (in pb) for exclusive $J/\psi \to \mu^+\mu^-$ photoproduction in $pp$ collisions, for different values of the c.m.s. energy $\sqrt{s}$ and different cuts on the muon pseudorapidities. Results are shown for the `bare' and `screened' cross sections, i.e. excluding and including soft survival effects, respectively, and the resulting average suppression due to these is also given. }
\label{table:photopsi}
\end{center}
\end{table}

In Table~\ref{table:photopsi} we show cross sections predictions for $J/\psi \to \mu^+ \mu^-$ production at $\sqrt{s}=7$ and 13 TeV, with the final--state muons restricted to lie within the LHCb acceptance ($2<\eta^\mu < 4.5$), as well as for central production ($-1<\eta^\mu < 1$) at $\sqrt{s}=13$ TeV. The muons are decayed including spin correlations, assuming $s$--channel helicity conservation in the $J/\psi$ production subprocess  and with the corresponding branching ratio taken from~\cite{Agashe:2014kda}. Predictions are shown for demonstration both with and without soft survival effects included, with in the latter case model 4 of~\cite{Khoze:2013dha} taken, although the results are in fact almost insensitive to this choice. This is to be expected: the main model dependence in the evaluation of the soft survival factor lies in the region of small impact parameter $b_t \ll R_p$, where $R_p$ is the proton radius, whereas the peripheral photoproduction process is relatively insensitive to this lower $b_t$ region.

The (screened) 7 TeV prediction is in excellent agreement with the LHCb measurement of~\cite{Aaij:2014iea}
\begin{equation}
\sigma^{J/\psi \to \mu^+\mu^-}(2<\eta^\mu < 4.5) = 291 \pm 7 \pm 19 \, {\rm pb}\;,
\end{equation}
where the first error is statistical and the second is systematic. However, it is important to emphasise that the predicted value depends sensitively on the precise form of the fit in (\ref{sigmaw}) to the $\gamma p \to J/\psi p$ subprocess cross section, in particular the value of the power $\delta_\psi$. As described above we have chosen a value for this which is at the lower end of the uncertainty band of the HERA fit. Taking a larger value will lead to an increase in the predicted cross section, with for example $\delta_\psi=0.70$, on the upper end of the uncertainty band, giving a $\sim 40 \%$ larger result, although a more precise evaluation of the uncertainty must account for the error on $N_\psi$ and the anti--correlation between this and $\delta_\psi$. We therefore choose this value to give a good fit to the LHCb data. However, this should be considered as a lower bound on the predicted cross sections, due to the low choice of $\delta_\psi$. It is therefore clear from Table~\ref{table:photopsi} that without the inclusion of soft survival effects, the LHCb data are in strong tension with such a fit to HERA data.

\begin{figure}
\begin{center}
\includegraphics[scale=0.6]{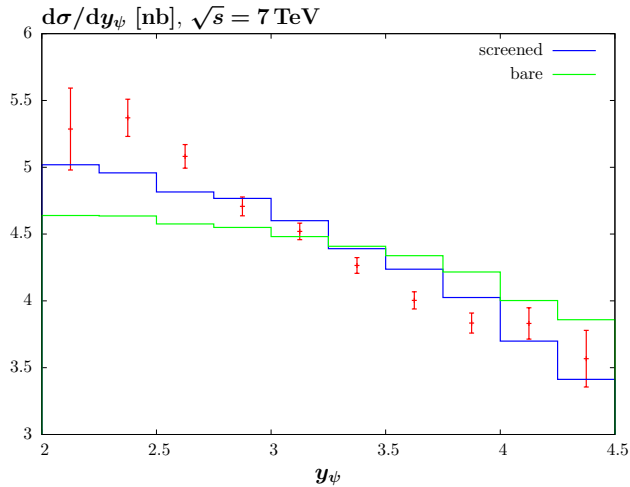}
\caption{Distributions with respect the $J/\psi$ rapidity $y_\psi$ at $\sqrt{s}=7$ TeV, compared to the LHCb data points from~\cite{Aaij:2014iea}. Theory curves corresponding to the `bare' and `screened' cross sections, i.e. excluding and including soft survival effects, respectively, are shown, and  the integrated cross sections are normalised to the data for display purposes. The correlated systematic errors are not shown.}
\label{fig:ypsi}
\end{center}
\end{figure} 

To examine the influence of survival effects further, we can also consider the distribution with respect to the $J/\psi$ rapidity, shown\footnote{The data points corrected from the fiducial measurement are shown so as to remove the influence of the muon cuts, giving a clearer demonstration of the underlying theory; as the correction factors are in fact derived in~\cite{Aaij:2014iea} using a previous version of \texttt{SuperChic}, these do not imply any significant model dependence.}  in Fig.~\ref{fig:ypsi}. As discussed in Section~\ref{sec:survqcd} the survival factor is not constant, and will therefore have an effect on the predicted distributions of the final--state particles. This is seen clearly in the figure, with the inclusion of screening corrections leading to a steeper fall--off with increasing rapidity. This is to be expected: as $y_\psi$ increases, so does the fractional momentum $\xi = M_\psi e^{y_\psi}/\sqrt{s}$ (for the dominant case that the photon is emitted from the proton moving in the positive $z$ direction), leading to a larger minimum photon $Q^2$, see (\ref{qi}). The reaction therefore becomes less peripheral, and the survival factor will decrease. This effect is also seen in Table~\ref{table:photopsi}, when comparing the average survival factor between the central and forward predictions. On the other hand, adjusting the input value of $\delta_\psi$ in (\ref{sigmaw}) within the range consistent with the HERA data leads to much smaller changes in the predicted distribution. Although the agreement is still far from perfect\footnote{However, we note that the correlated systematic errors are not shown in Fig.~\ref{fig:ypsi}.}, the overall trend of the data clearly prefer the screened prediction. While this conclusion is only strictly true in the context of the simple power--law HERA fit (\ref{sigmaw}), nonetheless this illustrates the importance of a full inclusion of soft survival effects in  theoretical models such as e.g.~\cite{Motyka:2008ac,Jones:2013pga}.

\begin{figure}
\begin{center}
\includegraphics[scale=0.6]{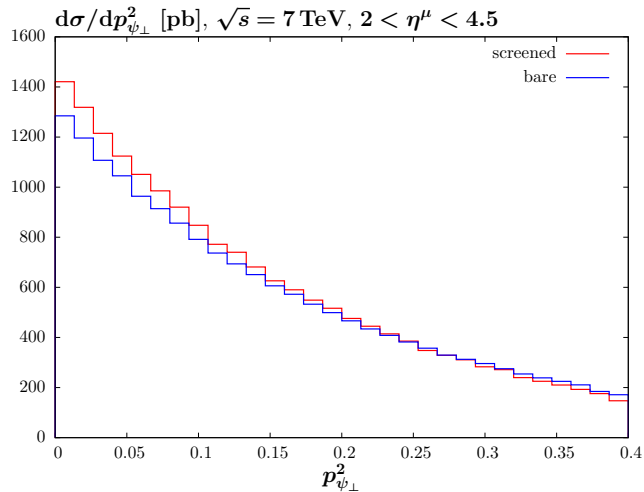}
\caption{Distributions with respect the $J/\psi$ transverse momentum at $\sqrt{s}=7$ TeV, corresponding to the `bare' and `screened' cross sections for $J/\psi\to \mu^+\mu^-$ production, i.e. excluding and including soft survival effects, respectively. The muons are required to have pseudorapidity $2<\eta^\mu < 4.5$, and the integrated bare cross section is normalised to the screened value for display purposes.}
\label{fig:ptpsi}
\end{center}
\end{figure} 

We may also consider the distribution with respect to the $J/\psi$ transverse momentum $p_{\psi_\perp}$. This is an important variable in the LHCb measurements~\cite{Aaij:2013jxj,Aaij:2014iea}, for which the selected events contain a non--negligible fraction with proton dissociation and/or additional particle production that falls outside the LHCb rapidity coverage. To subtract this background the measured $p_{\psi_\perp}^2$ distribution is fitted by a sum of two exponentials $\sim \exp (-b p_{\psi_\perp}^2)$, corresponding to the elastic and proton dissociative contributions. The data are well fit by such a parametric form, and in~\cite{Aaij:2014iea} LHCb find
\begin{equation}\label{bel}
b_{\rm el}^\psi = 5.70 \pm 0.11\,{\rm GeV}^{-2}
\end{equation}
for the pure elastic CEP contribution, while for the proton dissociative contribution the value of the corresponding slope is significantly smaller, reflecting the larger average $p_\perp$ in this case. Recalling that in CEP the vector sum of the proton transverse momenta is transferred directly to the produced object, this fitted value reflects a non--trivial interplay between the elastic electromagnetic and Pomeron form factors, given in (\ref{WWflux}) and (\ref{sigmaw}) respectively. While photon exchange generally prefers smaller values of the proton $p_\perp$ and so will have a smaller impact on $b_{\rm el}$ (\ref{bel}), this contribution cannot necessarily be neglected completely, in particular in the forward region where the slope of the Pomeron form factor in (\ref{bpsi}) can be quite high, and the average photon virtuality is larger. Again, we will expect screening corrections to have some influence on this value: in particular, as the expected suppression is larger at higher proton $p_\perp$, we will expect these to increase $b_{\rm el}$ compared to the bare case, see~\cite{HarlandLang:2009qe} for additional discussion. Such an effect, although fairly small, is clearly seen in Fig.~\ref{fig:ptpsi}, where the $J/\psi$ transverse momentum distributions in the screened and bare cases are shown. Performing a least--squares fit for $p_{\psi_\perp}^2<0.4\, {\rm GeV}^2$ we find the distributions can be well fitted by a simple exponential with slopes
\begin{equation}
b_{\rm el}^{\rm bare}=5.0\,{\rm GeV}^{-2}\qquad\qquad b_{\rm el}^{\rm sc.}=5.5\,{\rm GeV}^{-2}
\end{equation}
with a $\sim \pm\,  0.1 \,{\rm GeV}^{-2}$ error due to the uncertainty on the HERA fit~\cite{Aktas:2005xu} in (\ref{bpsi}), and a smaller error $\sim \pm\,  0.02 \,{\rm GeV}^2$  due to the fitting procedure. We can see that the bare result is inconsistent with the quite precise LHCb measurement (\ref{bel}),  but that the introduction of survival effects greatly reduces this tension. While the predicted rapidity distributions in Fig.~\ref{fig:ypsi} and the preference for screening corrections found in that case depend on the validity of the power--law fit (\ref{sigmaw}) outside the original $W_{\gamma p}$ region of the HERA fit, the parameterisation (\ref{bpsi}) is grounded in more fundamental principles of Regge theory: the value of the slope is driven by the structure of the Pomeron--proton vertex and the slope $\alpha'$ of the exchanged Pomeron, while the contribution from the heavy vector boson vertex will be very small. This behaviour is therefore expected to be present in more sophisticated models such as~\cite{Motyka:2008ac,Jones:2013pga}, and thus this result provides a more certain, and less model--dependent indication of the importance of a correct, fully differential, inclusion of survival effects. Interestingly, it appears that the predicted value may be somewhat lower than the measurement; further theoretical investigation of the model dependence of the result, as well as experimentally a more precise measurement of $b_{\rm el}$, in particular as a function of the $J/\psi$ rapidity, would help to clarify this. 

\begin{table}[h]
\begin{center}
\renewcommand\arraystretch{1.15}
\begin{tabular}{|c|c|c|c|}
\cline{2-4}
\multicolumn{1}{}{}&\multicolumn{2}{|c|}{$2<\eta^\mu<4.5$}&\multicolumn{1}{|c|}{$-1<\eta^\mu<1$}\\
\cline{2-4}
 \multicolumn{1}{c|}{}& $\sqrt{s}=7$ + 8 TeV &  $\sqrt{s}=13$ TeV&  $\sqrt{s}=13$ TeV \\ \hline
$\sigma^{\Upsilon}_{\rm scr.} $& 0.23 & 0.34& 0.29 \\
\hline
\end{tabular}
\caption{Cross section predictions (in pb) for exclusive $\Upsilon \to \mu^+\mu^-$ photoproduction in $pp$ collisions, including screening effects, for different values of the c.m.s. energy $\sqrt{s}$ and different cuts on the muon pseudorapidities. The 7 + 8 TeV result is given by the weighted average of the predictions corresponding to the relative fractions of integrated luminosity collected by the LHCb measurement~\cite{Aaij:2015kea} at these energies.}
\label{table:photoups}
\end{center}
\end{table}

Finally, considering the case of exclusive $\Upsilon(1S)\to \mu^+\mu^-$ production, this has recently been measured for the first time in hadronic collisions by the LHCb collaboration~\cite{Aaij:2015kea}, at $\sqrt{s}=7$ and $8$ TeV. As with $J/\psi$ production (\ref{sigmaw}) we take a power--law fit to the $\gamma \Upsilon$ cross section, but in this case the available HERA data~\cite{Adloff:2000vm,Chekanov:2009zz} are much less precise, and the resulting fit parameters are not well determined. The default fit in earlier versions of \texttt{SuperChic}, which was qualitatively similar to the LO prediction of~\cite{Jones:2013pga}, leads to much too steep an energy dependence in comparison to the LHCb data, and a new fit is therefore required. This is however complicated by the fact that in $pp$ collisions the photon may be emitted from either colliding proton, i.e. we must add the cross sections at energies $W^{\pm}=(M_\Upsilon \sqrt{s} e^{\pm y_\Upsilon})^{1/2}$; the lower energy $W^-$ contribution, due to the positive power $\delta$ in (\ref{sigmaw}), is expected to be smaller, but not necessarily negligible. A simultaneous fit to these contributions must therefore be performed, and as the effect of changing $\delta$ typically acts in opposite directions on the $W^{\pm}$ contributions, this leads to a large amount of cancelation between these, such that a good fit to the combined cross section can be achieved for a wide range of $\delta$, in particular for the current fairly limited LHCb and HERA data. This highlights a wider issue: when attempting to extract information about the $\gamma p\to Vp$ cross section from measurements in $pp$ collisions some additional assumptions must be made in separating the $W^+$ and $W^-$ contributions. A more robust extraction may be made by increasing the statistics and $W$ range of the data, but ultimately a direct and model independent comparison in $pp$ collisions can only be performed against lab frame variables, such as the meson rapidity distribution, where both $W^+$ and $W^-$ contributions are suitably included in the theory prediction. On the other hand, in $pA$ collisions this issue does not arise, as the source of photons can to very good approximation be uniquely identified with the heavy ion, due to the $Z^2$ enhancement in the photon flux.

Assuming a simple power--law as in (\ref{sigmaw}) and fitting to the available data leads to a quite low best fit value of $\delta\approx 0.3$, but with a sizeable error. Such a low value is disfavoured on general grounds~\cite{Newman:2013ada}, from which we would expect a larger $\delta$ than in the case of the lower scale $J/\psi$ production process. Moreover, due to the cancellation effects discussed above, it is possible to achieve a good fit for much higher values of $\delta$, and instead we set $\delta_\Upsilon=0.7$, with the normalisation $N_\Upsilon=5.7$ pb  found from the resulting fit, for which we still have $\chi^2/{\rm d.o.f}\sim 1$. The weighted average of the predicted cross sections within the LHCb acceptance at $\sqrt{s}=7$ and 8 TeV, corresponding to the relative fractions of integrated luminosity collected by LHCb~\cite{Aaij:2015kea} at these energies, is shown in Table~\ref{table:photoups}, and is seen to be reasonably consistent with the measurement of $0.22\pm 0.07$ pb. Predictions for $\sqrt{s}=13$ TeV are also given: these contain a $\sim 50\%$ uncertainty due to the error in the extracted parameters of the power--law fit.

\subsection{Two--photon mediated processes}\label{sec:2photo}
\begin{figure}
\begin{center}
\includegraphics[scale=0.6]{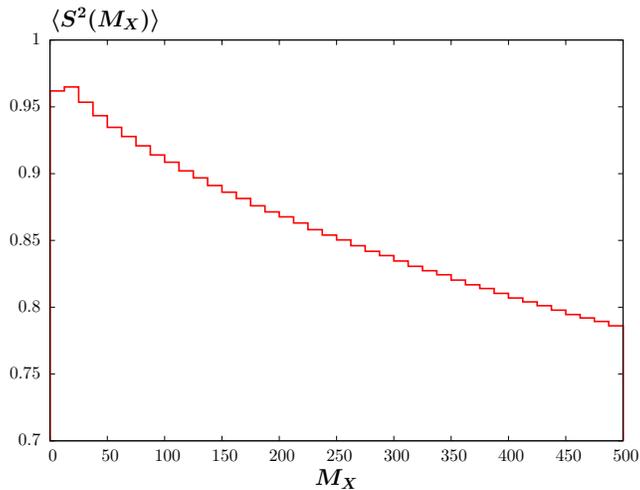}
\caption{Average survival factor $\langle S^2_{\rm elk} \rangle={\rm d}\sigma_{\rm scr.}/{\rm d}\sigma_{\rm bare}$ as a function of the central system invariant mass $M_X$ for muon pair production, at $\sqrt{s}=14$ TeV. The muons are required to have $p_\perp> 2.5$ GeV and $|\eta|<2.5$.}
\label{fig:wwsurv}
\end{center}
\end{figure}

\begin{table}[h]
\begin{center}
\renewcommand\arraystretch{1.15}
\begin{tabular}{|c|c|c|c|c|}
\hline
&$\mu^+\mu^-$ &$\mu^+\mu^-$, $M_{\mu\mu}>2 M_W$&$\mu^+\mu^-$, $p_\perp^{\rm prot.}<0.1$ GeV&$W^+W^-$\\
\hline
$\sigma_{\rm bare} $& 6240 & 11.2& 3170& 87.5 \\
$\sigma_{\rm sc.} $ & 5990 & 9.58&3150 & 71.9 \\
$\langle S^2_{\rm eik}\rangle$ & 0.96 & 0.86&0.994 &0.82 \\
\hline
\end{tabular}
\caption{Cross section predictions (in fb) for exclusive muon and $W$ boson pair production at $\sqrt{s}=13$ TeV. The muons are required to have  $p_\perp> 5$ GeV and $|\eta|<2.5$, and are shown with and without an additional cut of $M_{\mu\mu}>2 M_W$, while in the $W$ boson case, no cuts are imposed. Results are shown for the `bare' and `screened' cross sections, i.e. excluding and including soft survival effects, respectively, and the resulting average suppression due to these is also given.}
\label{table:llww}
\end{center}
\end{table}
\noindent In this section we present a very brief selection of results for the two-photon exclusive production of lepton (electron and muon) and $W$ boson pair production. In Table~\ref{table:llww} we show predictions for the muon and $W$ boson pair production cross sections, with and without soft survival effects included. In the case of muon pair production we can see that, as expected from the discussion in Section~\ref{sec:photo}, the average soft suppression factor is close to unity, due to the peripheral two--photon interaction, as well as the vanishing of the $T_{\pm \pm}$ amplitudes for massless leptons discussed in Section~\ref{sec:photoscreen}. However, as seen in the previous section, as the system invariant mass increases, we will expect the photon momentum fraction $x_\gamma \propto M_X$ to increase. This will lead to a higher average photon virtuality, see (\ref{qi}), and therefore for the average survival factor to be smaller for this less peripheral interaction. We also show the prediction for the same muon pair cross section, but subject to the requirement that $M_{\mu \mu}> 2 M_{W}$; while the suppression factor is still quite close to unity, it is clearly lower. This reduction in the survival factor with $M_X$ is seen more clearly in Fig.~\ref{fig:wwsurv} where the average suppression is shown for muon pair production as a function of the pair invariant mass; a very similar result is found in the case of $W$ pair production. We also show in Table~\ref{table:llww} the total $W$ boson pair production cross section, where the suppression factor is smaller still, due to the different helicity structure of the production amplitudes (for which the $T_{\pm\pm}$ amplitudes are non--vanishing).  Finally, the muon pair production cross section, but with the outgoing protons required to have transverse momentum $p_\perp<0.1$ GeV is shown: by placing such a cut, the reaction is required to be highly peripheral, and it can be seen that the suppression factor is extremely close to unity. On the other hand, as discussed in~\cite{HarlandLang:2012qz}, in the case that one or both protons dissociate the reaction is generally much less peripheral, and a proper inclusion of soft survival effects becomes crucial; this can lead to sizeable deviations in the data with respect to the result of e.g. the LPAIR MC~\cite{Vermaseren:1982cz,Baranov:1991yq}, which does not include these effects. 

\begin{table}
\begin{center}
\renewcommand\arraystretch{1.15}
\begin{tabular}{|c|c|c|}
\hline
&$\mu^+\mu^-$ &$e^+e^-$\\
\hline
$\sigma_{\rm bare} $& 0.795 & 0.497 \\
$\sigma_{\rm sc.} $ & 0.742 & 0.459\\
$\langle S^2_{\rm eik}\rangle$ & 0.93 & 0.92\\
\hline
ATLAS data~\cite{Aad:2015bwa}&$0.628\pm 0.032\pm 0.021$&$0.428\pm 0.035\pm 0.018$\\
\hline
\end{tabular}
\caption{Cross section predictions (in pb) for exclusive muon and electron pair production at $\sqrt{s}=7$ TeV. The muons (electrons) are required to have  $p_\perp> 10(12)$ GeV, and in both cases $|\eta^l|<2.4$. Results are shown for the `bare' and `screened' cross sections, i.e. excluding and including soft survival effects, respectively, and the resulting average suppression due to these is also given. These are compared to the ATLAS data~\cite{Aad:2015bwa}.}
\label{table:llATLAS}
\end{center}
\end{table}

Recently, the ATLAS collaboration have published a measurement of exclusive $\mu^+\mu^-$ and $e^+e^-$ production~\cite{Aad:2015bwa} in normal LHC running conditions, by vetoing on additional charged--particle tracks associated with the lepton vertex, and applying further corrections to extract the exclusive signal. This is compared to the MC predictions in Table~\ref{table:llATLAS}. The bare cross sections are in both cases too high compared to the data, but a better agreement is achieved when survival effects are included. However, interestingly, while there is excellent agreement within uncertainties in the electron case, the prediction for the muon cross section lies $\sim$ 3 $\sigma$ above the data, i.e. a lower value of the average soft suppression appears to be preferred. Such a discrepancy may indicate that a further refinement of the modelling of the opacity in the high $b_t$ region, to which two--photon induced processes are sensitive, is required, or alternatively may be a result of contamination from non--exclusive events due, for example, to proton dissociation, although a detailed attempt is made in~\cite{Aad:2015bwa} to subtract this background and account for any uncertainty on this in the systematic error on the data. Further measurements, ideally differential in $m_{ll}$, as well as with tagged protons, thus effectively eliminating the possibility of proton dissociation, will be of great use in clarifying this issue. It is worth emphasising that as the two--photon production process is theoretically so well understood, this represents a particularly clean probe of soft survival effects, in particular if the outgoing protons are tagged.

 These results highlight the importance of a proper treatment of screening corrections, which is still often not included in the literature. In the recent work of~\cite{daSilveira:2015hha} for example, where the question of constraining the photon PDF in exclusive $l^+l^-$ and $W^+W^-$ production is considered, soft survival effects, which as noted above may be particularly important  if proton dissociative events are included, are omitted\footnote{Moreover, in the semi--exclusive case it is not the standard photon PDF which enters in the hard cross section. Rather, the PDF must be evolved using a modified form of the DGLAP equation in which emission in the experimentally relevant rapidity region is forbidden. This will be the subject of a future study~\cite{HLfut}.}. Another important example of this is in~\cite{Dyndal:2014yea}, where an evaluation of the survival factor for two--photon induced processes is given, the predictions of which are compared to in the ATLAS data in~\cite{Aad:2015bwa}. While a differential treatment of the survival factor is given, and for example the same qualitative decrease with $M_X$ as in Fig.~\ref{fig:wwsurv} is seen, the correct photon $q_\perp$ dependence, described in Section~\ref{sec:photoscreen}, is not included in this work; in impact parameter space, only the $b_t$ dependence of the photon flux (\ref{WWflux}) is included, and not that of the $\gamma\gamma \to X$ subprocess. This omits entirely any process--dependence in the survival factor, and will not give a reliable estimate for the expected suppression. After an explicit calculation, we find that including only the $b_t$ dependence of the photon flux, as in~\cite{Dyndal:2014yea}, tends to underestimate the survival factor by $\sim 10-20$ \% for lepton pair production within the ATLAS event selection. It should be emphasised that this does not correspond to a genuine model dependence: the correct inclusion of the photon $q_\perp$ dependence as discussed in Section~\ref{sec:photoscreen} follows simply from the derivation of the equivalent photon approximation and cannot be omitted. Indeed, in the case of lepton pair production, if we include the correct photon $q_\perp$ dependence at the amplitude level, and instead of using the model of~\cite{Khoze:2013dha} we take the simplified form for the proton opacity used in~\cite{Dyndal:2014yea}, then the predicted survival factor is almost unchanged for these peripheral interactions.

\subsection{Heavy quarkonia production}\label{sec:quark}
The CEP of $\chi_{c,b}$ (and $\eta_{c,b}$) quarkonia states has been considered in~\cite{HarlandLang:2009qe,HarlandLang:2010ep,HarlandLang:2010ys}. These processes are implemented in the new MC, and in this section we present updated predictions for a short selection of $\chi_{c,b}$ cross sections. However, we note that a number of theoretical updates and modifications have been included in comparison to these earlier studies: we summarise these below first.

As discussed in these earlier studies, the predicted cross section for $\chi_c$ CEP has a significant theoretical uncertainty. One important source of this is the gluon PDF, which at the quite low $x$ and $Q^2$ values relevant to the process is not well determined; for this reason, in~\cite{HarlandLang:2009qe,HarlandLang:2010ep}, consistently with the earlier treatment in~\cite{Khoze:2004yb}, the $\chi_c$ cross section was calculated at a lower value of $\sqrt{s}=60$ GeV, where the gluon is better constrained, and a Regge extrapolation $\sigma \propto s^{2\Delta}$, with $\Delta=0.2$, was assumed to calculate the cross section at higher energies. While the prediction of such an approach was observed to give good agreement with the CDF measurement of $\chi_{cJ}$ production~\cite{Aaltonen:2009kg}, clearly it depends on the validity of the Regge scaling assumption, and in particular on the precise value of $\Delta$. Such a scaling closely corresponds to assuming a typical power--like low--$x$ form $xg\sim x^{-\Delta}$ in the gluon PDF, which appears in the perturbative amplitude (\ref{bt}) via (\ref{fgskew}). However, while typically gluon PDFs extracted from global fits exhibit to good approximation such a behaviour, the precise value of $\Delta=\partial \log xg/\partial \log x$, which will depend on the PDF set, may be different from $\Delta=0.2$ and, crucially, will depend on the scale $Q^2$, due to DGLAP evolution; such physics will be missed by the simple assumption above.

More physically, we now therefore treat quarkonium production consistently with other exclusive processes within the MC, and evaluate the cross section at the appropriate $\sqrt{s}$ value, without any explicit Regge scaling assumption. As the CDF measurement of $\gamma\gamma$ CEP~\cite{Aaltonen:2011hi} strongly favours the results from typical LO gluon PDFs, see e.g.~\cite{Harland-Lang:2014efa}, a finding which is  further supported by the LHCb measurement of $J/\psi$ pair production at $\sqrt{s}=7$ and 8 TeV~\cite{Aaij:2014rms}, these should be the appropriate choice here. However, the low $x$ dependence of such PDFs leads to a significantly steeper $\sqrt{s}$ dependence: for example, with MMHT14LO PDFs~\cite{Harland-Lang:2014zoa} for a typical $Q^2$  given by the average gluon transverse momentum $\langle Q_\perp^2 \rangle \sim 4$ ${\rm GeV}^2$ we find $\Delta \approx 0.32$. The effect of instead taking such a value of $\Delta$ is dramatic, leading to a factor of $\sim$ 5 larger cross section at Tevatron energies, and $\sim$ an order of magnitude larger at the LHC. The result of explicitly including the corresponding PDF set in (\ref{bt}) within the MC is consistent with this increase, with the precise energy dependence being somewhat steeper still. This leads to a predicted total $\chi_{cJ}$ cross section at $\sqrt{s}=1.96$ TeV that is significantly larger than the value measured by CDF~\cite{Aaltonen:2009kg}.

There are however other corrections to the simplified approach of~\cite{HarlandLang:2009qe,HarlandLang:2010ep} which we may consider. In particular, in these works, the $\chi_{c,b}$ cross sections are given in terms of the $gg \to \chi$ vertices
\begin{align}\label{V0}
&V_{0^+}=\sqrt{\frac{1}{6}}\frac{c_\chi}{M_\chi}(3M_\chi^2(q_{1_{\perp}}q_{2_{\perp}})-(q_{1_{\perp}}q_{2_{\perp}})(
q_{1_{\perp}}^2+q_{2_{\perp}}^2)-2q_{1_{\perp}}^2q_{2_{\perp}}^2)  \; ,\\ \label{V1}
&V_{1^+}=-\frac{2ic_\chi}{s} p_{1,\nu}p_{2,\alpha}((q_{2_\perp})_\mu(q_{1_\perp})^2-(q_{1_\perp})_\mu(q_{2_\perp})^2)\epsilon^{\mu\nu\alpha\beta}\epsilon^{*\chi}_\beta  \; ,\\
\label{V2}
&V_{2^+}=\frac{\sqrt{2}c_\chi M_\chi}{s}(s(q_{1_\perp})_\mu(q_{2_\perp})_\alpha+2(q_{1_\perp}q_{2_\perp})p_{1\mu}p_{2\alpha})\epsilon_\chi^{*\mu\alpha}  \; ,
\end{align}
where $q_{i\perp}$ is the transverse momentum of the incoming gluon $i$, with $q_{i\perp}^2=-{\bf q}_{i\perp}^2$, and
\begin{equation}\label{cchi}
c_\chi=\frac{1}{2\sqrt{N_C}}\frac{16 \pi \alpha_S}{(q_1 q_2)^2}\sqrt{\frac{6}{4\pi M_\chi}}\phi'_P(0)\;.
\end{equation}
Crucially, in~\cite{HarlandLang:2009qe,HarlandLang:2010ep} the vertices (\ref{V0})--(\ref{V2}) were calculated in the ${\bf q}_{i\perp}^2\ll M_\chi^2$ limit.  
This approximation was justified by the fact that, while a more complete treatment of the particle kinematics, as in  the $k_\perp$--factorisation approach of the Durham model, includes an important part of the higher--order corrections (within collinear factorisation) which generate initial--state gluon off--shellness, the $q_{i\perp} \sim M_\chi$ region may be particularly sensitive to the exact form of such corrections, for which a full NLO (and beyond) calculation would be necessary to give a complete evaluation. Moreover, the derivation of (\ref{bt}) is only strictly valid in the ${\bf q}_{i\perp}^2\ll M_\chi^2$ regime. 

On the other hand, as discussed in~\cite{HarlandLang:2010ys}, see also~\cite{Pasechnik:2007hm}, keeping the full $q_{i\perp}$ dependence in $gg\to \chi$ amplitudes can have a significant effect, and it may still be the case that this gives a more reliable estimate of the expected cross sections. Moreover, as we will now show, such a treatment is preferred by the existing data on $\chi_c$ CEP, when the LO PDFs described above are used. For this reason, we now choose to include the full gluon $q_{i\perp}$ dependence everywhere in the vertices (\ref{V0})--(\ref{V2}). The most significant effect of this can be seen by observing that the term appearing in the denominator of (\ref{cchi})
\begin{equation}\label{q1q2}
(q_1 q_2)=\frac{1}{2}\left(M_\chi^2-q_{1\perp}^2-q_{2\perp}^2\right)\;,
\end{equation}
which in the on--shell gluon approximation is equal to $M_\chi^2/2$, appears to the fourth power in the cross section, so that even for relatively small average $\langle {\bf q}_{i\perp}^2\rangle/M_\chi^2$, the correction may be sizeable. Moreover, in general this ratio is not too small: using the vertex (\ref{V0}) in (\ref{bt}), for example, we have $\langle {\bf q}_{i\perp}^2\rangle/M_\chi^2\sim 0.3$. As this will lead to a larger denominator (\ref{q1q2}), the effect is to reduce the expected cross section. The precise value of this depends on such factors as the object mass, PDF choice, and $\sqrt{s}$, but also on the particle spin, through the form of the vertex (\ref{V0})--(\ref{V2}), all of which affect $\langle {\bf q}_{i\perp}^2\rangle$. Taking $\sqrt{s}=60$ GeV, so as to compare with the predictions of~\cite{HarlandLang:2009qe,HarlandLang:2010ep}, we find that including the full $q_{i\perp}$ dependence in the $gg\to \chi$ amplitudes  reduces the expected cross sections for central $\chi_{cJ}$ production by factors of $\sim$ 3, 6 and 2 for the $J=0,1$ and $2$ states, respectively. The reduction in the cross sections is sizeable, and the spin dependence is clear, while at larger $\sqrt{s}$ the higher average gluon $Q_\perp^2$ leads to a further increase in the suppression. For the $\chi_{c1}$, the impact is particularly severe: we recall that the production amplitude vanishes for on--shell incoming gluons and in the forward proton limit, and therefore we would expect the predicted cross section to be particularly sensitive to the precise form of the gluon off--shell corrections, as well as the proton $p_\perp$ spectra. In the case of the higher mass $\chi_b$ the effect is greatly reduced, leading to a factor of $\sim$ 10--20\% smaller cross section for the $\chi_{b0}$ and $\chi_{b2}$ and a factor of $\sim 2$ smaller cross section for the $\chi_{b1}$.

We find that this overall reduction in the expected cross sections is largely cancelled by the increase due to the use of the LO PDFs at the appropriate $\sqrt{s}$ value, so that the final predicted cross sections at the Tevatron are now in good agreement with the CDF measurement~\cite{Aaltonen:2009kg} of $\sum \sigma(\chi_{cJ}\to J/\psi\gamma)=0.97$ nb, see Table~\ref{table:chictev} below. As discussed above, this PDF choice is well motivated by the agreement it gives with the CDF measurement of $\gamma\gamma$ CEP~\cite{Aaltonen:2011hi} and the LHCb measurement of $J/\psi$ pair production~\cite{Aaij:2014rms}, as well as being theoretically simpler and arguably more justified then invoking additional Regge scaling arguments. The full inclusion of off-shell gluon effects then appears to be required to give good agreement with the CDF $\chi_c$ data~\cite{Aaltonen:2009kg} when such LO PDFs are used. It should however be emphasised that while the agreement with these data is equally good between this updated approach and that of~\cite{HarlandLang:2009qe,HarlandLang:2010ep}, the predictions for other observables are quite distinct. This different treatment of the gluon off-shellness affects the $\chi_{c1}$ and $\chi_{c2}$ to $\chi_{c0}$ ratios, and the predicted energy dependence of the $\chi_c$ cross sections is steeper, leading to increased rates at the LHC. In addition, as the higher mass $\chi_b$ states are much less sensitive to off--shell gluon effects we predict larger cross sections for these, see Table~\ref{table:chiblhcb} below.

A further change compared to the earlier works relates to the normalisation of the derivative of the wave function at the origin $\phi'_P(0)$ in (\ref{cchi}). As in~\cite{HarlandLang:2010ep} this is normalized in terms of the total $\chi_{c0}$ width
\begin{equation}\label{gamchi}
\Gamma_{\rm tot}(\chi_{0})\approx \Gamma(\chi_0\to gg)=96\frac{\alpha_S^2 }{M_{\chi_0}^4}|\phi'_P(0)|^2\;.
\end{equation}
This implicitly includes higher order contributions to the $gg \to\chi$ vertex, under the assumption that these are the same as in the $\chi \to gg$ case. While in~\cite{HarlandLang:2010ep}  an additional subtraction was made so that these corrections were only included for the spin--zero $\chi_0$, by dividing the extracted value of $|\phi'_P(0)|^2$ for the $\chi_{1,2}$  by a K--factor, taken to be 1.5, we now choose to apply this correction uniformly; in this case the ratio of cross sections between the spin states correspond to the purely LO results, consistently with the rest of the calculation. This change leads to a factor of 1.5 smaller $\chi_{c0}$ cross section compared to previous results. In the case of the $\chi_b$, the measurement of~\cite{Kornicer:2010cb} suggests that earlier estimates for the total width, such as that taken in~\cite{HarlandLang:2010ys}, underestimate the true value, and prefers $\Gamma_{\rm tot}(\chi_{b0})\approx 0.8$ MeV, excluding any NLO K--factor. We take this value here: as in~\cite{HarlandLang:2010ys} this was taken as the value including a K-factor (again assumed to be 1.5), the $\chi_{b0}$ predictions are unchanged, while the $\chi_{b(1,2)}$ predictions are now a factor of 1.5 larger.

Finally, while in~\cite{HarlandLang:2009qe,HarlandLang:2010ep} an additional `non--perturbative' contribution was considered in the case of $\chi_c$ production, this is no longer included. This does not rule out the possibility of important non--perturbative effects for the relatively low mass $\chi_c$ states; rather, we choose to consistently only consider the perturbative mechanism in the MC. The influence of possible non--perturbative effects can then be seen as deviations from these predictions, for example in the ratio of cross sections between the different spin states. In addition, as discussed above, we recall that no enhanced survival effects are considered here: these may reduce the expected cross sections somewhat, but will leave the ratios $\chi_{1}/\chi_{0}$ etc almost unchanged.

\begin{table}[h]
\begin{center}
\renewcommand\arraystretch{1.15}
\begin{tabular}{|c|c|c|c|c|c|}
\hline
$\sigma\times {\rm Br}$&$\chi_{c0}$ &$\chi_{c1}$&$\chi_{c2}$&$\sum \chi_{cJ}$&$\frac{\chi_{c0}}{\sum \chi_{cJ}}$\\
\hline
CTEQ6L1&0.29&0.12&0.22&0.63&0.46\\
\hline
MMHT14LO&0.38&0.11&0.29&0.78&0.49\\
\hline
NNPDF3.0&0.51&0.15&0.41&1.1&0.48\\
\hline
\end{tabular}
\caption{Differential cross section, ${\rm d}\sigma/{\rm d} y_\chi$, predictions (in nb) at $y_\chi=0$ for exclusive $\chi_{cJ} \to J/\psi\gamma$ production at $\sqrt{s}=1.96$ TeV, including screening effects. The sum over the three spin states, and the ratio of the $\chi_{c0}$ to the total cross sections are also given. Results are shown for three choices of LO PDF: CTEQ6L1~\cite{Pumplin:2002vw}, MMHT14~\cite{Harland-Lang:2014zoa}, NNPDF 3.0 ($\alpha_S(M_Z^2)=0.130$)~\cite{Ball:2014uwa}, and with model 1 of~\cite{Khoze:2013dha} for the soft survival factor.}
\label{table:chictev}
\end{center}
\end{table}

\begin{table}[h]
\begin{center}
\renewcommand\arraystretch{1.15}
\begin{tabular}{|c|c|c|c|c|c|}
\hline
$\sigma\times {\rm Br}$&$\chi_{c0}$ &$\chi_{c1}$&$\chi_{c2}$&$\frac{\chi_{c1}}{\chi_{c0}}$&$\frac{\chi_{c2}}{\chi_{c0}}$\\
\hline
CTEQ6L1&19&8.3&15&0.44&0.79\\
\hline
MMHT14LO&51&17&41&0.33&0.80\\
\hline
NNPDF3.0&23&7.1&19&0.31&0.83\\
\hline
\hline
LHCb data &$9.3\pm 4.5$&$16.4\pm7.1$&$28.0\pm12.3$&$1.8\pm 0.7$&$3.0\pm 0.9$\\
\hline
\end{tabular}
\caption{Cross section predictions (in pb) for exclusive $\chi_{cJ} \to J/\psi\gamma\to \mu^+\mu^- \gamma$ production at $\sqrt{s}=7$ TeV, including screening effects, with the cross section ratios between the different spin states shown for clarity. Results are shown for three choices of LO PDF: CTEQ6L1~\cite{Pumplin:2002vw}, MMHT14~\cite{Harland-Lang:2014zoa}, NNPDF 3.0 ($\alpha_S(M_Z^2)=0.130$)~\cite{Ball:2014uwa}, with model 1 of~\cite{Khoze:2013dha} for the soft survival factor, and with the $\chi_c$ decay products in the rapidity region $2<\eta<4.5$. The preliminary LHCb measurement~\cite{LHCb} is shown for comparison, with statistical, systematic and luminosity errors added in quadrature; for the cross section ratios only the statistical errors are included.}
\label{table:chiclhcb}
\end{center}
\end{table}

Turning now to the MC predictions, in Table~\ref{table:chictev} we show the predicted differential cross sections at $y_\chi=0$ for $\chi_{cJ}$ production at the Tevatron for a range of PDF sets, and with model 1 of the soft survival factor from~\cite{Khoze:2013dha}. This can be compared to the previous predictions of~\cite{HarlandLang:2010ep}. As described above, the steeper energy dependence due to the LO gluon PDF sets and the suppression due to the new treatment of off--shell gluon effects are found to largely cancel, such that the total $\chi_{cJ}$ cross section is similar in size, and in all cases consistent with the CDF measurement~\cite{Aaltonen:2009kg}.  On the other hand the ratios $\chi_{c(1,2)}/\chi_{c0}$ are different. From the discussion above, in the case of the $\chi_{c1}/\chi_{c0}$ ratio, we expect the updated off--shell gluon effects to reduce this by a factor of $\sim 3$, while the $\chi_{c2}/\chi_{c0}$ ratio should be increased by a factor of $\sim 2$, and in both cases the new treatment of the NLO K--factor for the $\chi_{c0}$ will increase these by a factor of $1.5$. We therefore expect the $\chi_{c1}/\chi_{c0}$ ratio to be decreased by a factor of $\sim 2$ and the  $\chi_{c2}/\chi_{c0}$ ratio to be increased by a factor of $\sim 3$: this is indeed consistent with the results of Table~\ref{table:chictev}. It should be emphasised that as the changes due to the updated treatment of the off--shell gluon effects are theoretically motived, and essential to reproduce the CDF $\chi_c$ data~\cite{Aaltonen:2009kg} with the LO PDFs sets preferred by measurements of exclusive processes at the Tevatron and LHC~\cite{Aaltonen:2009kg,Aaij:2014rms} (as well as themselves being a consistent choice for the LO theoretical calculation), these may be expected to better describe the data; the difference due to the treatment of the NLO K--factor, on the other hand, which enters at the level of the uncertainty due to higher--order corrections, simply corresponds to one possible prescription, with everything treated at LO, while the approach~\cite{HarlandLang:2010ep}, where some approximate K--factors are included, may give a better estimate. Finally, we note that the ratio $\chi_{c0}/\sum \chi_{cJ}$, also shown in Table~\ref{table:chictev}, is reasonably consistent with the more recent CDF limit for this of $\lesssim 50\%$, calculated from the corresponding limits on $\chi_{c0}$ production in the $\pi^+\pi^-$ and $K^+K^-$ channels~\cite{Aaltonen:2015uva}. 

\begin{table}[h]
\begin{center}
\renewcommand\arraystretch{1.15}
\begin{tabular}{|c|c|c|c|c|c|}
\hline
$\sigma\times {\rm Br}$&$\chi_{c0}$ &$\chi_{c1}$&$\chi_{c2}$&$\frac{\chi_{c1}}{\chi_{c0}}$&$\frac{\chi_{c2}}{\chi_{c0}}$\\
\hline
CTEQ6L1&30&14&26&0.46&0.87\\
\hline
MMHT14LO&77&24&68&0.31&0.88\\
\hline
NNPDF3.0&39&11&35&0.28&0.90\\
\hline
\end{tabular}
\caption{Cross section predictions (in pb) for exclusive $\chi_{cJ} \to J/\psi\gamma\to \mu^+\mu^- \gamma$ production at $\sqrt{s}=13$ TeV, including screening effects, with the cross section ratios between the different spin states shown for clarity. Results are shown for three choices of LO PDF: CTEQ6L1~\cite{Pumplin:2002vw}, MMHT14~\cite{Harland-Lang:2014zoa}, NNPDF 3.0 ($\alpha_S(M_Z^2)=0.130$)~\cite{Ball:2014uwa},  with model 1 of~\cite{Khoze:2013dha} for the soft survival factor, and with the $\chi_c$ decay products in the rapidity region $2<\eta<4.5$.}
\label{table:chiclhcb13}
\end{center}
\end{table}

In Table~\ref{table:chiclhcb} we show predictions for $\chi_c\to J/\psi\gamma\to\mu^+\mu^-\gamma$ CEP at $\sqrt{s}=7$ TeV for the same event selection as the preliminary LHCb measurement~\cite{LHCb} of this process. While the default predictions with the MMHT14LO PDFs are somewhat too large, the agreement is better, although not perfect, with the CTEQ6L and NNPDF3.0 PDF sets. These are on the other hand only the predictions for the central PDFs: for the MMHT14LO set, for example, the PDF uncertainty is found to be $\pm \sim 50\%$. The $\chi_{c(1,2)}/\chi_{c0}$ ratios also exhibit some generally small PDF dependence, which is larger in the $\chi_{c1}$ case. These ratios are qualitatively consistent with the predictions, and it should be emphasised that this is in itself a highly non--trivial result: for example, for the $\chi_{c2}$ without the effect of the $J_z=0$ selection rule we would expect the corresponding ratio to be $\sim 2$ orders of magnitude higher, as is observed inclusively. However there remains some indication of tension with the LHCb data in both cross section ratios, which, as discussed in detail in~\cite{HarlandLang:2010ep}, may be due to the influence of proton dissociation on the selected events, but may also indicate that further theoretical work is needed to better model the CEP of these relatively low mass states. Further measurements of this, with the new HERSCHEL detectors~\cite{Albrow:2014lta} at LHCb allowing a much greater rejection of non--exclusive events, will greatly clarify this question; the corresponding cross section predictions at $\sqrt{s}=13$ TeV are shown in Table~\ref{table:chiclhcb13}.

A further observable, which is not expected to suffer from the same uncertainties as in the $\chi_c$ case, is the CEP of the heavier $\chi_b$ states: for example, as discussed above, the effect of adjusting the treatment of gluon off--shellness is much smaller, due to the larger meson mass, and more generally the production process is safely in the perturbative regime. In Table~\ref{table:chiblhcb} we show predictions for $\chi_{bJ}\to \Upsilon\gamma\to\mu^+\mu^-\gamma$ production at $\sqrt{s}=13$ TeV, within the LHCb acceptance. Due to the larger mass, the predicted cross sections are significantly smaller, although still experimentally realistic. The expected $\chi_{b1}/\chi_{b0}$ ratio is much smaller than in the $\chi_c$ case, due to the higher meson mass, while the $\chi_{b2}/\chi_{b0}$ ratio is somewhat smaller, mainly due to the differing branching ratios for the particle decays. We note that the predicted cross sections are significantly larger than the estimates of~\cite{HarlandLang:2010ep}: the \texttt{SuperChic} v1.47 MC predicts for example a $\chi_{b0}$ cross section of $8$ fb within the same acceptance. This sizeable difference is primarily due to the steeper energy dependence induced by the LO PDF sets.

\begin{table}[h]
\begin{center}
\renewcommand\arraystretch{1.15}
\begin{tabular}{|c|c|c|c|c|c|}
\hline
$\sigma\times {\rm Br}$&$\chi_{b0}$ &$\chi_{b1}$&$\chi_{b2}$&$\frac{\chi_{b1}}{\chi_{b0}}$&$\frac{\chi_{b2}}{\chi_{b0}}$\\
\hline
CTEQ6L1&73&5.3&27&0.073&0.37\\
\hline
MMHT14LO&110&6.6&43&0.060&0.39\\
\hline
NNPDF3.0&62&3.5&24&0.056&0.39\\
\hline
\end{tabular}
\caption{Cross section predictions (in fb) for exclusive $\chi_{bJ} \to \Upsilon\gamma\to \mu^+\mu^- \gamma$ production at $\sqrt{s}=13$ TeV, including screening effects, with the cross section ratios between the different spin states shown. Results are shown for three choices of LO PDF: CTEQ6L1~\cite{Pumplin:2002vw}, MMHT14~\cite{Harland-Lang:2014zoa}, NNPDF 3.0 ($\alpha_S(M_Z^2)=0.130$)~\cite{Ball:2014uwa},  with model 1 of~\cite{Khoze:2013dha} for the soft survival factor, and with the $\chi_b$ decay products in the rapidity region $2<\eta<4.5$.}
\label{table:chiblhcb}
\end{center}
\end{table}

Finally, it is interesting to observe the dependence of the average survival factor on the spin on the produced states. This is shown in Table~\ref{table:chicsurv} for $\chi_c$ production at $\sqrt{s}=1.96$ and 7 TeV, and for two different soft models in~\cite{Khoze:2013dha}: the average suppression is much weaker for the higher spin states, due to the more peripheral nature of the interaction vertices (\ref{V1}) and (\ref{V2}), while the stronger overall reduction as the collider energy is increased, and the variation between the model choices, is clear.

\begin{table}[h]
\begin{center}
\renewcommand\arraystretch{1.15}
\begin{tabular}{|c|c|c|c|c|c|c|}
\cline{2-7}
\multicolumn{1}{}{}&\multicolumn{3}{|c|}{Model 1}&\multicolumn{3}{|c|}{Model 4}\\
\cline{2-7}
\multicolumn{1}{c|}{}&$\chi_{c0}$ &$\chi_{c1}$&$\chi_{c2}$&$\chi_{c0}$ &$\chi_{c1}$&$\chi_{c2}$\\
\hline
$\langle S^2_{\rm eik}\rangle$, Tevatron&0.027&0.067&0.060&0.033&0.088&0.077\\
\hline
$\langle S^2_{\rm eik}\rangle$, LHCb&0.014&0.029&0.036&0.019&0.041&0.050\\
\hline
\end{tabular}
\caption{Average survival factors for $\chi_{cJ}$ production corresponding to Tables~\ref{table:chictev} (`Tevatron') and ~\ref{table:chiclhcb} (`LHCb'), for two different model choices of the soft survival factor defined in~\cite{Khoze:2013dha}. Values are shown for MMHT14 LO~\cite{Harland-Lang:2014zoa} PDFs, but the results for other choices are similar.}
\label{table:chicsurv}
\end{center}
\end{table}

\subsection{Generated processes: summary}\label{sec:other}
In the coming years the LHC will take increasingly precise data at unprecedented energies. This presents the possibility for a wide programme of exclusive measurements, building on those already performed during Run--I~\cite{yp}. For this reason a wide range of processes, some of which have been discussed in detail above, are included in the MC. We summarise these, and give some motivations for further experimental measurements, below (see~\cite{HarlandLang:2009qe,HarlandLang:2010ep,Harland-Lang:2013ncy,Harland-Lang:2014efa} for further details of the processes generated):

\begin{itemize}

\item Exclusive jet and vector meson production, discussed further in Sections~\ref{sec:excjets} and~\ref{sec:exvec}, respectively. 

\item $\chi_{cJ}$ production, discussed in Section~\ref{sec:quark}, by both the $J/\psi \gamma \to \mu^+\mu^- \gamma$ and two--body decay channels, including full spin correlations. As discussed in Section~\ref{sec:quark}, the existing preliminary LHCb measurement~\cite{LHCb} of exclusive $\chi_c\to J/\psi \gamma$ production, while qualitatively agreeing with the Durham model expectations, shows some possible tension with the theory, in particular with an apparent excess of $\chi_{c2}$ events visible.
Further measurements of this process, with the new HERSCHEL detectors~\cite{Albrow:2014lta} at LHCb allowing a much greater rejection of non--exclusive events, will greatly clarify this question. Observables such as two body decays, where the $\chi_{c2}$ is no longer enhanced by the $J/\psi \gamma$ branching (and the $\chi_{c1}$ will be absent entirely), as well as more differential observables such as the $\chi_{c(1,2)}/\chi_{c0}$ ratio as a function of $p_\perp(\chi)$ will allow a closer comparison to theory.

\item Light meson pair ($\pi\pi$, $KK$, $\eta(')\eta(')$, $\phi\phi$) production. Measurements of the meson $p_\perp$ distribution are of particular importance, as these are sensitive to the transition from the `non--perturbative' low $p_\perp$ region where a Regge--based model may be more appropriate\footnote{Such a model is not included in \texttt{SuperChic}: see~\cite{Harland-Lang:2013dia,Lebiedowicz:2015eka} for MC implementations.} to the higher $p_\perp$ region where the perturbative approach described in~\cite{HarlandLang:2011qd} should become relevant. Where such a transition occurs is currently a very uncertain question. However, we note that the perturbative approach makes very distinct predictions about, for example, the cross section for flavour--non--singlet meson pair production, compared to Regge--based expectations.

\item $\gamma\gamma$ production. As well as being of interest in its own right, this relatively clean process can serve as a useful benchmark to compare with other CEP measurements: by considering cross section ratios with respect to $\gamma\gamma$ CEP, uncertainties due to the gluon PDFs and survival factor largely cancel. In addition, as discussed in~\cite{d'Enterria:2013yra} the light--by--light scattering process has yet to be seen experimentally, and may represent an observable at the LHC, or in a future circular collider; the  photon--induced $\gamma\gamma\to \gamma\gamma$ process is therefore also included.

\item Double $J/\psi$ and $\psi(2S)$ quarkonia production. In~\cite{Aaij:2014rms} LHCb have reported the observation of double $J/\psi$ and $J/\psi \psi(2S)$ production, in broad agreement with theory expectations in~\cite{Harland-Lang:2014efa}. More precise measurements, in particular in the case of $\psi(2S)J/\psi$ and double $\psi(2S)$ production (so far not seen), as well as more differential measurements of the double $J/\psi$ cross sections, for example with respect to the meson rapidity separation and $p_\perp$, would provide a useful test of the perturbative approach. As discussed above, a measurement of the ratio to other exclusive processes such as $\gamma\gamma$ would also greatly reduce the potentially large theoretical uncertainties. In addition, although not currently included in the MC, the cross section for double $\eta_c$ production could be of the same size or even larger than for the $J/\psi$, due to the particular Feynman diagrams which enter in this case. 

\item $\chi_b$ production, by both the $\Upsilon \gamma \to \mu^+\mu^- \gamma$ and two--body decay channels, including full spin correlations. The higher mass compared to the $\chi_c$ case means that the process should be safely perturbative, so that we should expect good agreement between the measured $\chi_{b2}/\chi_{b0}$ ratio and theory, while the $\chi_{b1}$ should be practically absent. 

\item Pseudoscalar $\eta_{c,b}$ meson production. These are expected to be strongly suppressed by the $J^P_z=0^+$ selection rule. A measurement of these processes would therefore provide a direct test of this; for the $\eta_c$, the predicted cross sections are large enough that this could represent a realistic observable, although it may be challenging to identify an experimentally viable decay channel.

\item $W^+W^-$ and $l^+l^-$ production ($l=e,\mu, \tau$), via two--photon collisions, see Section~\ref{sec:2photo}. This purely QED cross section is known theoretically to very high precision, and in the case of lepton pair production, if cuts are place to restrict the proton $p_\perp$ (or equivalently $p_\perp(l^-l^+)$) to low values the survival factor may be essentially omitted, so that such a process could be used as a luminosity calibration~\cite{Khoze:2000db}. Alternatively, without placing such cuts, this well understood process can serve as a particularly clean probe of soft survival effects, in particular in the presence of tagged protons.
 Exclusive $W^+W^-$ production is of particular interest as a probe of potential anomalous gauge couplings, see~\cite{Kepka:2008yx,Chapon:2009hh,Royon:2015coa}. All photon--induced processes are also available for $e^+e^-$ initial states.

\item SM Higgs boson production. The CEP of a SM Higgs, in precisely the mass region observed by ATLAS and CMS~\cite{Aad:2012tfa,Chatrchyan:2013lba}, has received a lot of attention. As discussed further in~\cite{Harland-Lang:2014lxa}, the exclusive mode is very well suited to probe crucial identification issues such as the $b\overline{b}$ coupling and the $CP$--parity of this object. 

\item The photoproduction of $\rho(770)$ and $\phi(1020)$ mesons, via the $\pi^+\pi^-$ and $K^+K^-$ decay channels, respectively. As for other photoproduction processes, the simple fits (\ref{sigmaw}) and (\ref{bpsi}) for the $W$ dependence of the cross section and slope parameter $b$ are taken. The fit of~\cite{Breitweg:1999jy} to the effective Pomeron trajectories is used, and the normalisations are set using data from~\cite{Derrick:1996af,Breitweg:1997ed}. These give $\delta=0.19$ (0.16) and $\alpha^\prime= 0.125$ (0.158) ${\rm GeV}^{-2}$, with $\sigma=11.4$ (0.96) $\mu {\rm b}$ at $W_{\gamma p}= 75$ (70) GeV and $b=11.1$ (7.3) ${\rm GeV}^{-2}$ at $W_{\gamma p}= 84$ (70) GeV,  for the $\rho$ ($\phi$). 

\end{itemize}

In addition, although not currently included in the MC, the measurement of `exotic'  quarkonia in the exclusive channel may provide further information about these states. For example, the $X(3915)$, for which there has been recent discussion as to whether this can be interpreted as a $\chi_{c0}(2P)$ state~\cite{Chen:2013yxa,Olsen:2014maa}, could be searched for exclusively in the $D\overline{D}$ channel. The $Z(3930)$, currently associated with the $\chi_{c2}(2P)$, may also be searched for via the $D\overline{D}$ mode; as mentioned in~\cite{HarlandLang:2010ys}, the expected cross section for such an excited state should be similar to the $\chi_{c2}(1P)$. In addition, as discussed further in~\cite{Harland-Lang:2014lxa}, a clear observation of exclusive $X(3872)$ production can provide important constraints on its nature. Other possibilities include the $J/\psi\phi$ narrow resonances, such as the $X(4350)$, clearly seen by Belle~\cite{Shen:2009vs}, for which $0^{++}$ or $2^{++}$ assignments, as well as tetraquark or more conventional charmonia interpretations, remain possible. Such a resonance peak could be searched for in the exclusive channel, as part of more general resonance searches in for instance the $\psi'\phi$ channel. A further process not currently included in the MC is exclusive $D\overline{D}$ production, for which quite large cross sections are expected, and is sensitive to the $c\overline{c}\to D\overline{D}$ transition with no additional particle production. This latter process is the subject of a future work~\cite{HLfut}, including a full MC implementation.

\section{\texttt{SuperChic 2}: code and availability}\label{sec:MC}

\texttt{SuperChic 2} is a Fortran based Monte Carlo that can generate the processes described in Section~\ref{sec:other}, with and without soft survival effects. User--defined distributions may be output, as well as unweighted events in the HEPEVT and Les Houches formats. The code and a user manual can be found at \texttt{http://projects.hepforge.org/superchic}.

\section{Summary and outlook}\label{conc}

In this paper we have presented results from a new Monte Carlo for central exclusive production (CEP): \texttt{SuperChic 2}. In a CEP process, an object $X$ is produced, separated by two large rapidity gaps from intact outgoing protons, with no additional hadronic activity. Theoretically, the study of CEP requires the development of a framework which is quite different from that used to describe the inclusive processes more commonly considered at hadron colliders, while experimentally it results in a very clean signal (in the absence of pile up), and the outgoing protons can be reconstructed by proton tagging detectors, situated far down the beam line. 

\texttt{SuperChic 2} builds on the earlier \texttt{SuperChic} event generator, but includes a range of theoretical updates and improvements. Most significant of these relates to the soft survival factor, representing the probability of no additional underlying event activity, which would spoil the exclusivity of the final state. In previous MC implementations, this has simply been treated as a constant quantity. However, this is not in general true, and we have demonstrated in this paper a number of cases where such an assumption breaks down. For the first time, \texttt{SuperChic 2}  includes a fully differential treatment of the soft survival factor, and we have seen that this can lead to some distinct, and model--dependent, predictions for the corresponding particle distributions.

In addition, a much wider range of processes has been included, with exclusive 2 and 3 jet production being a particularly relevant addition; this is the first inclusion of the latter process in a publicly available MC. Other production processes currently included are: vector meson ($\rho,\phi,J/\psi,\psi(2S),\Upsilon(1S)$)  photoproduction, $\chi_{c,b}$ and $\eta_{c,b}$ quarkonia, photon pairs, light meson pairs, heavy ($J/\psi$, $\psi(2S)$) quarkonia pairs, SM Higgs boson and two--photon initiated lepton, $W$ boson and photon pair production (with both $pp$ and $e^+e^-$ initial states).

We have described the theoretical approach implemented in the \texttt{SuperChic 2} MC, and have considered in more detail a selection of the processes listed above: in Section~\ref{sec:excjets} we have considered 2 and 3--jet production, and discussed the possibility for observing the `radiation zeros' described in~\cite{Harland-Lang:2015faa} in the 3 jet case; in Section~\ref{sec:exvec} we have considered $J/\psi$ and $\Upsilon(1S)$ photoproduction, showing how a correct inclusion of soft survival effects leads to important differences in the predicted meson rapidity and transverse momentum distributions; in Section~\ref{sec:2photo} we have considered two--photon initiated muon and $W$ boson pair production, and shown how the predicted survival factor depends on the system mass $M_X$; in Section~\ref{sec:quark} we have described an updated approach to modelling quarkonium CEP, and presented results for $\chi_{c,b}$ production.

In light of the measurement possibilities for exclusive processes during Run--II of the LHC~\cite{yp}, the study of CEP processes is very topical: exclusive events may be measured with both protons tagged using the approved and installed AFP~\cite{CERN-LHCC-2011-012} and CT--PPS~\cite{Albrow:1753795} forward proton spectrometers, associated with the ATLAS and CMS central detectors, respectively, see also~\cite{Royon:2015tfa}, as well as using rapidity gap vetoes to select dominantly exclusive event samples, for example at LHCb, where the relatively low instantaneous luminosity and wide rapidity coverage allowed by the newly installed HERSCHEL forward detectors~\cite{Albrow:2014lta} are highly favourable. It is our aim that the new  \texttt{SuperChic 2} MC will play an important role in this promising physics programme.
 
\section*{Acknowledgements}

We thank David d'Enterria, Phillip Harris, Daniel Johnson, Oldrich Kepka, Ronan Mcnulty, Risto Orava, Matthias Saimpert and Gustavo Gil Da Silveira for useful discussions. LHL thanks the Science and Technology Facilities Council (STFC) for support via the grant award ST/L000377/1. MGR thanks the IPPP at the University of Durham for hospitality. The work by MGR was supported by the RSCF Grant  14-22-00281. VAK thanks the Leverhulme Trust for an Emeritus Fellowship.

\appendix

\section{Jet production: helicity amplitudes}\label{app:jets}
In this appendix we give some further details about the contributing amplitudes in the exclusive jet cross section. The helicity amplitudes relevant to quark dijet production are
\begin{align}\label{qqexcjz0}
\mathcal{M}_{\pm\pm,h\bar{h}} &=\frac{\delta^{cd}}{2 N_c}\frac{16\pi\alpha_s}{(1-\beta^2\cos^2\theta)}\frac{m_q}{M_X}(\beta h \pm 1)\delta_{h,\bar{h}}\;,\\ \label{qqexcjz2}
\mathcal{M}_{\pm\mp,h\bar{h}}&=\frac{\delta^{cd}}{2 N_c}\frac{8\pi\alpha_s}{(1-\beta^2\cos^2\theta)}\beta h \sin \theta\left(2\delta_{h,\bar{h}} \frac{m_q}{M_X}\sin\theta \mp \delta_{h,-\bar{h}}(1\mp h\cos\theta)\right)\;,
\end{align}
while for the gluon amplitudes we have
\begin{align}\label{ggexcjz0}
\mathcal{M}_{\pm\pm,\pm\pm}&=\delta^{CD} \frac{N_c}{N_c^2-1} \frac{32\pi\alpha_s}{(1-\cos^2\theta)}\;,\\ \label{ggexcjz2}
\mathcal{M}_{\pm\mp,+-} &=\mathcal{M}_{\mp\pm,-+}=\delta^{CD} \frac{N_c}{N_c^2-1} 8\pi\alpha_s\left( \frac{1\pm \cos\theta}{1\mp\cos\theta}\right)\;,
\end{align}
for gluons of `$\pm$'  helicity and quarks of helicity $h/2$, while $c,d$ are the outgoing quark color labels, and $\beta=(1-4m_q^2/M_X^2)^{1/2}$. As discussed in~\cite{Harland-Lang:2014efa}, these amplitudes are shown for the azimuthal angle $\phi=0$ for the outgoing particle momenta. In general, some helicity amplitudes have an overall $\phi$--dependent phase, which while having no effect in the spin--summed inclusive cross section, must be included when using the decomposition (\ref{Agen}) to give the correct result in the exclusive case.

The amplitudes relevant to 3--jet production are all `maximally helicity violating' (MHV)~\cite{Mangano:1990by}, and can therefore be written down quite simply, see~\cite{Harland-Lang:2015faa} for further details.

\bibliography{references}{}
\bibliographystyle{h-physrev}

\end{document}